\documentclass[11pt,a4paper]{article}
\usepackage{jcappub}

\usepackage{graphicx}	
\graphicspath{{./}{../}}
\usepackage{amsmath}	
\usepackage{empheq}
\usepackage{xcolor}

\usepackage{multirow}
\usepackage{booktabs}
\usepackage{makecell}
\usepackage{caption}  

\usepackage{CJKutf8}
\usepackage{bm}

\newcommand{\be}{\begin{equation}}
\newcommand{\ee}{\end{equation}}
\newcommand{\bea}{\begin{eqnarray}}
\newcommand{\eea}{\end{eqnarray}}
\def\ba#1\ea{\begin{align}#1\end{align}}

\renewcommand{\refeq}[1]{Equation (\ref{eq:#1})}

\def\({\left(}
\def\){\right)}
\def\[{\left[}
\def\]{\right]}
\def\<{\left<}
\def\>{\right>}

\def\ln{{\rm ln}}
\def\max{{\rm max}}

\definecolor{RedWine}{rgb}{0.743,0,0}

\definecolor{NavyBlue}{rgb}{0,0,0.9}

\title
{Lossless Compression of Cosmological Information from Type~Ia Supernova Distance Measurements}

\author[1]{Zhenyuan Wang (\begin{CJK*}{UTF8}{gbsn}王震远\end{CJK*})}
\author[1]{, Yun Wang}

\affiliation[1]{IPAC, California Institute of Technology, 1200 E. California Blvd., Pasadena, CA 91125, U.S.A}

\emailAdd{zywang@ipac.caltech.edu}
\emailAdd{wang@ipac.caltech.edu}

\abstract{%
We perform model-independent distance measurements on four Type~Ia supernovae (SNe~Ia) compilations (Pantheon, Pantheon+, DES-Dovekie, Union3) and compress each dataset into the values of $\log r_p(z)$ at eleven redshift knots, where $r_p(z)$ is a rescaled comoving distance. These Gaussian distributed compressed values, together with their full covariance, completely capture the distance-redshift relation information from each dataset. We demonstrate this by using these to perform an Markov Chain Monte Carlo (MCMC) likelihood analysis to infer cosmological parameters in flat $\Lambda$CDM, flat $w_0 w_a$CDM, and a non-parametric reconstruction of the dark-energy density $X(z) \equiv \rho_{\rm DE}(z)/\rho_{\rm DE}(0)$.
Across all datasets and flux-averaging configurations and all three cosmological models, the resulting parameter contours and figures of merit reproduce the corresponding full distance-modulus analyses using the original SNe~Ia data sets (see Wang \& Wang 2026a \cite{PaperI}) within the statistical sampling noise of the chains, demonstrating that the eleven $\log r_p$ data points are an operationally lossless compression of the cosmological information in the dataset.
Our SN Ia data compression enables an analytic analysis that
completes in $\mathcal{O}(10^{-2})$\,s per dataset and reduces the downstream cosmological MCMC to the fast evaluation of an $11$-dimensional Gaussian likelihood, with a per-step cost set by the number of knots and independent of the SNe~Ia sample size.
Our methodology will benefit the data analysis of future surveys from Euclid, Roman, and LSST, which will deliver SNe~Ia samples one to three orders of magnitude larger than current ones.
}


\begin{document}
\maketitle


\section{Introduction}
\label{sec:intro}

Type Ia supernovae (SNe~Ia) gave the first direct evidence that the expansion of the Universe is accelerating \cite{Riess1998, Perlmutter1999}. The current generation of compilations (Pantheon+ \cite{Scolnic2022}, DES-Dovekie \cite{DESY5_2024, Dovekie2025}, Union3 \cite{Rubin2023}) each contain $\mathcal{O}(10^3)$ SNe Ia. Combined with DESI~DR2 BAO and CMB, these data give the current best constraints on the dark energy equation of state $w(z)$ \cite{DESIDR2, DESIDR2_DE}. These SNe data (except Union3) are released as per-SN distance moduli with an $N \times N$ covariance. That format works for a single likelihood evaluation, but it does not show the distance--redshift shape each survey prefers, and it is not portable between surveys. Any reanalysis under a new dark energy model requires running the full Markov Chain Monte Carlo (MCMC) again. Whether dark energy is consistent with a cosmological constant is still an open question, and the answer will come from the next generation of SNe~Ia surveys.

The next decade will see SNe~Ia samples increase by factors of roughly $10$ from Roman and Euclid, and up to $\sim 10^3$ from LSST. The Euclid mission targets $\sim 1.8 \times 10^4$ SNe Ia out to $z \sim 1.5$ \cite{Astier2014}; the Roman High-Latitude Time-Domain Survey (HLTDS) is designed to yield $\sim 10^4$ SNe~Ia out to $z \sim 2$ \cite{RomanSNe, Rose2021, HounsellHLTDS2025}; and the Vera C.\ Rubin Observatory Legacy Survey of Space and Time (LSST) will discover $\sim 10^5$--$10^6$ well-measured SNe~Ia over its ten-year operation \cite{LSSTSciBook}. The per-SN analysis approach will not scale to those large sample sizes. Covariance matrices grow as $N^2$, joint likelihoods across surveys and rapid model exploration become more computationally challenging. We need a compact, model-independent intermediate data product: a simple Gaussian distance prior compressing the SN Ia data that any downstream cosmological analyses can use directly, analogous to the CMB distance priors $(R, l_a, \omega_b)$ that compress the entire $C_\ell$ spectrum into three numbers \cite{WangMukherjee2007, WangDai2016}. We use the term \emph{lossless} in an operational sense throughout this paper. We require that the downstream cosmological posterior from the compressed data agree with the posterior from the full $(\bm{\mu}, \mathbf{C}_\mu)$ both in the Figure-of-Merit, $\mathrm{FoM} \equiv 1/\sqrt{\det \mathbf{C}_{\bm\theta}}$ with $\mathbf{C}_{\bm\theta}$ the covariance matrix of the cosmological parameters $\bm\theta$ \citep{Wang2008}, and in the 1D and 2D marginalized posteriors. The FoM check alone is necessary but not sufficient, since the determinant could in principle preserve the total posterior volume while redistributing variance across parameter directions, which would change the 1D marginal posterior and the correlation structure.

The framework we build on is the rescaled distance $r_p(z) \equiv [H_0 r(z)/(cz)] (1+z)^{\mathcal{W}(z)}$ of \cite{Wang2009, ZhaiWang2019}, where $\mathcal{W}(z)$ is an empirical exponent of redshift chosen to keep $r_p(z) \approx 1$ across the observed range. \cite{Wang2009, ZhaiWang2019} took the constant choice $\mathcal{W}(z) = 0.41$, which keeps $r_p$ approximately flat over $0 < z \lesssim 1.5$. The values of $r_p$ at a set of redshift knots are treated as free parameters. A cubic spline through the knots gives $r_p(z)$ at any redshift, which determines $d_L(z)$ and hence the predicted distance modulus $\mu_{\rm th}(z)$ at each SN's redshift; the knot values are then constrained by matching $\mu_{\rm th}$ to $\mu_{\rm obs}$ via MCMC. This is model-independent by construction: knot values are set by the data, and the spline only does local interpolation between them. We stay with the cubic spline rather than switching to a Gaussian process \cite{Holsclaw2010, Shafieloo2012} or Chebyshev polynomials \cite{DESIDR2_DE} because the spline does not require a kernel or assuming any global features. Local interpolation between knots is the minimum needed to capture the features of the data.

We make three changes relative to \cite{Wang2009, ZhaiWang2019}. The first is the rescaling exponent. Replacing the constant $0.41$ with a quadratic function in $u$, where $u = z/(1+z)$, makes the rescaled $r_p(z)$ smooth enough across $0 < z < 3$ that a small number of cubic-spline knots reproduce its shape accurately. The second is the choice of variable. Working in $\log r_p$ rather than $r_p$ is what makes a closed-form solution available: under the standard SN~Ia likelihood (distance modulus $\bm{\mu}_{\rm obs}$ with {\it Gaussian} covariance $\mathbf{C}_\mu$), our spline forward model is exactly linear in the $\log r_p$ knots, and the posterior on the knots is exactly Gaussian. Therefore, the mean and covariance of $\log r_p$ can be obtained by the best linear unbiased estimator (BLUE) which costs $\mathcal{O}(10^{-2})\,\mathrm{s}$ per dataset and avoids both the convergence diagnostics and the high-$z$ non-Gaussianity that shows up when the $r_p$ knots are sampled directly via MCMC (Sec.~\ref{subsec:validation}). 
The third is consistency: no new likelihood assumption is introduced beyond the standard $\mu$-Gaussian one, so the $\log r_p$ posterior is what the traditional $\bm\mu$-based analysis would give if it were inverted analytically. A similar approach at the catalog level is the Union3 distance-modulus compression of \cite{Rubin2023}, which uses a cubic spline through $22$ redshift knots to represent the residual $\mu - \mu_{\rm fid}(z)$ relative to a fiducial flat $\Lambda$CDM with $\Omega_m = 0.3$. Our $\log r_p$ representation is mathematically a linear reparameterization of the same spline-on-residual idea, since $\log r_p \propto \mu - \nu(z)$ with $\nu(z)$ a deterministic function of redshift. The $\log r_p$ knots retain only the data-driven shape of the distance--redshift relation. Relative to Union3 we (i)~adopt a uniform $11$-knot grid across all four compilations, and (ii)~use a closed-form BLUE estimator (Sec.~\ref{subsec:closed_form}), so the compression itself doesn't require MCMC. 

Flux averaging, introduced by \cite{Wang2000}, addresses a separate concern: the weak-lensing magnification PDF for high-$z$ SNe is non-Gaussian and skewed, so direct averaging of distance moduli is biased when individual events are scattered by lensing magnification. Averaging in flux space rather than in magnitude space removes the lensing-caused bias \cite{Wang2002, Wang2005}. In the traditional MCMC analysis, flux averaging is applied dynamically: at every sampled cosmology the 
SN Ia data set $(\bm\mu_{\rm obs}, \mathbf{C}_\mu)$ is re-binned using that sample's $d_L(z)$ to set the luminosity weights and bin redshifts. This data-cosmology coupling makes the flux-averaged forward model nonlinear in the $\log r_p$ knots: the conversion to flux from $\mu$ is exponential, and the luminosity-weighted bin average depends on the predicted $d_L^2(z)$ used to scale individual SNe. We handle this with an iterative approach, which performs the same dynamic re-binning as MCMC but in a deterministic loop: at each iteration the previous-best-fit $\log r_p$ knots are used to flux-average the per-SN data into binned distance moduli $(\bar{\bm\mu}_{\rm obs}, \bar{\mathbf{C}})$ on $\sim 4\times$ as many redshift bins as $\log r_p$ knots, and a closed-form BLUE step on the binned data then updates the $\log r_p$ knots. Convergence takes $2$--$4$ iterations, so the compression with flux averaging completes in seconds without MCMC.

In Paper~I \cite{PaperI}, we reported cosmological constraints from full MCMC on distance modulus data $\bm{\mu}$ for Pantheon, Pantheon+, DES-Dovekie, and Union3 in flat $\Lambda$CDM, $w_0 w_a$CDM, and a non-parametric model for the dark energy density ratio $X(z) \equiv \rho_{\rm DE}(z)/\rho_{\rm DE}(0)$. This paper provides the corresponding compressed data product for the same full datasets: a closed-form Gaussian posterior (mean and covariance) on $11$ $\log r_p$ knots that any downstream pipeline can use as a Gaussian likelihood, reproducing the Paper~I results accurately and orders of magnitude faster. 

The paper is organized as follows. Section~\ref{sec:method} sets up the framework: the $r_p(z)$ definition and rescaling (Sec.~\ref{subsec:rp_definition}), the spline forward model and likelihood (Sec.~\ref{subsec:spline_reconstruction}), the closed-form Gaussian posterior on $\log r_p$ (Sec.~\ref{subsec:closed_form}), 
the iterative-Newton extension for flux-averaged data (Sec.~\ref{subsec:fa_newton}), and a practical recipe summarizing the workflow of compression and downstream cosmological analysis(Sec.~\ref{subsec:recipe}). Section~\ref{sec:results} introduces the four SNe~Ia compilations, validates the closed-form/iterative-Newton estimators against MCMC (Sec.~\ref{subsec:validation}), presents the $\log r_p$ reconstructions, examines the empirical Gaussianity of the compressed posterior, verifies losslessness through downstream FoM ratios, and reports the cosmological constraints. We summarize this paper and discuss future applications in Section~\ref{sec:discussion}. Mathematical background, including the BLUE (Appendix~\ref{app:linear_gaussian}) and Newton's method (Appendix~\ref{app:newton_review}), is reviewed in Appendix~\ref{app:logrp_closed_form}. The compressed $\log r_p$ data products for all four SN Ia compilations are tabulated in Appendix~\ref{app:compressed_data}.

\section{Methodology: Compression Framework and Estimators}
\label{sec:method}

This section sets up the model-independent distance measurement as data compression that the rest of the paper builds on. We define the rescaled comoving distance $r_p(z)$ and the rescaling function that makes it smooth across the observed redshift range in Sec.~\ref{subsec:rp_definition}. Then we describe how to use interpolation to measure the $\log r_p$ at redshift knots from the distance modulus data $\bm{\mu}$ in Sec.~\ref{subsec:spline_reconstruction}. In Sec.~\ref{subsec:closed_form} we derived the closed-form best linear unbiased estimator (BLUE) for $\log r_p$ knots and their Gaussian covariance in the case without flux averaging. We then extend BLUE to an iterative-Newton estimator to include the flux averaging in Sec.~\ref{subsec:fa_newton}. In Sec.~\ref{subsec:recipe} we summarize with a practical recipe for compressing the distance modulus data into $\log r_p$ knots and Gaussian covariance matrix, and for using the compressed product in downstream cosmological analyses.

\subsection{Definition}
\label{subsec:rp_definition}

The rescaled distance was first introduced by \cite{Wang2009}, who proposed the power-law form $r_p(z) = [H_0 r(z)/(cz)]\,(1+z)^{0.41}$, later adopted in \cite{ZhaiWang2019}. Here $r(z)$ is the radial comoving distance. In a flat universe, $r(z)$ is given by
\begin{equation}
    r(z) = \frac{c}{H_0}\int_0^z \frac{dz'}{E(z')},\quad E(z) \equiv H(z)/H_0,
    \label{eq:comoving_distance}
\end{equation}
so the explicit $H_0$ in the $r_p$ prefactor cancels the $c/H_0$ in $r(z)$, and $r_p(z)$ depends only on the dimensionless expansion rate $E(z)$. The construction therefore encodes only the shape of the expansion history. The constant exponent $0.41$ was chosen empirically to keep $r_p \approx 1$ over $0 < z \lesssim 1.5$. We extend this construction by promoting the exponent to a quadratic function of $u \equiv z/(1+z)$:
\begin{equation}
    r_p(z) \equiv \frac{H_0 r(z)}{c z} (1+z)^{{\mathcal W}(z)},\quad {\rm where\;\;} {\mathcal W}(z) = 0.223 + 0.186\,u + 0.245\,u^2
    \label{eq:cadillac_transform}
\end{equation}
The function $\mathcal{W}(z)$ keeps $r_p(z)$ close to unity across $0 < z < 3$, so a small number of cubic-spline knots reproduce its shape accurately across the redshift range covered by current and future SNe~Ia compilations.

\begin{figure*}[t]
    \centering
    \includegraphics[width=\linewidth]{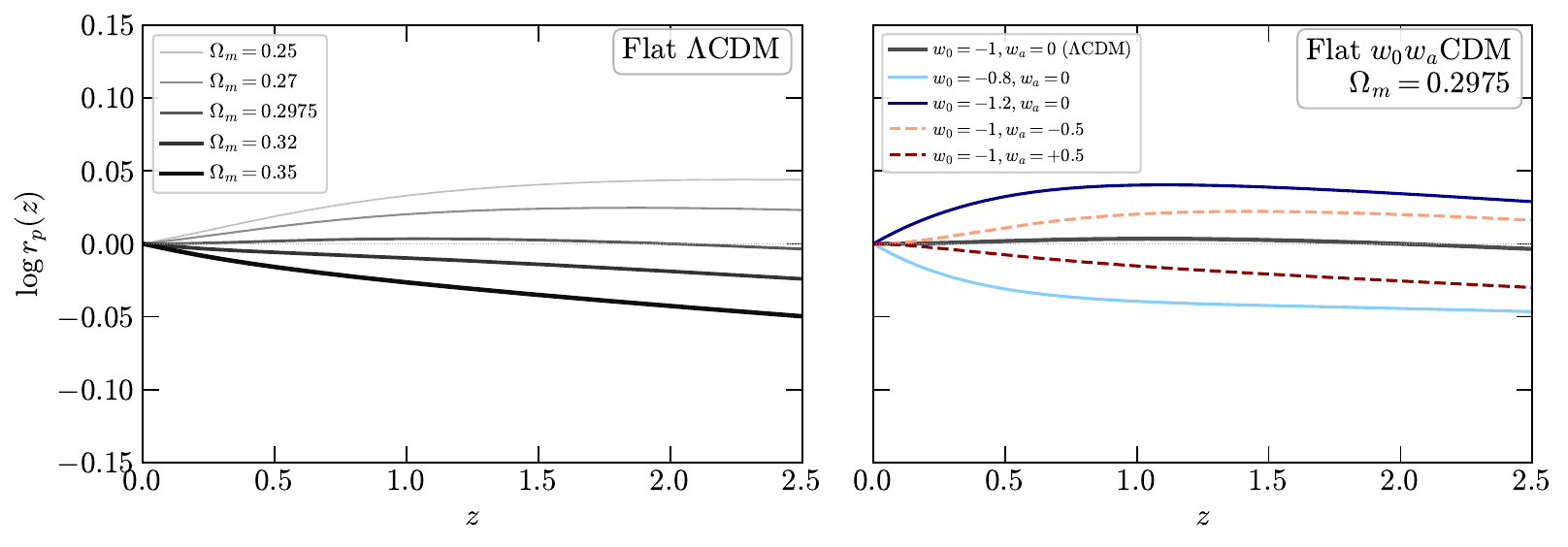}
    \caption{Theoretical $\log r_p(z)$ (natural log) for representative cosmology models, computed from the full distance integral with the rescaling exponent $\mathcal{W}(z) = 0.223 + 0.186\,u + 0.245\,u^2$. {\it Left:} flat $\Lambda$CDM with $\Omega_m \in \{0.25, 0.27, 0.2975, 0.32, 0.35\}$, plotted in a grey family with line darkness and thickness increasing with $\Omega_m$. {\it Right:} flat $w_0 w_a$CDM with $\Omega_m$ fixed at the fiducial value. Solid blue and cyan lines vary $w_0$ at $w_a = 0$, dashed orange and brown lines vary $w_a$ at $w_0 = -1$. The fiducial $\Lambda$CDM ($w_0 = -1$, $w_a = 0$, $\Omega_m=0.2975$) appears as the same medium-grey thick line in both panels.}
    \label{fig:cosmography_2panel}
\end{figure*}

Figure~\ref{fig:cosmography_2panel} demonstrates two complementary points. First, across all reasonable cosmologies $\log r_p(z)$ remains smooth across the full redshift range, confirming that the rescaling does its job and a small number of spline knots can capture the full shape (Sec.~\ref{subsec:spline_reconstruction}). Second, distinct cosmologies leave distinct smooth signatures on $\log r_p$, with $\Omega_m$ (left panel) and $w_0$, $w_a$ (right panel) each tracing a separable pattern across the displayed redshift range. The amplitude of these cosmological signatures ($\sim 0.05$ at $z \sim 1$) is comparable to the per-knot $1\sigma$ uncertainty achievable from current SNe~Ia compilations (Sec.~\ref{sec:results}), so measuring $\log r_p(z)$ knot by knot directly visualizes the distance--redshift shape the data prefer and the cosmological models that can reproduce it.

\subsection{Spline interpolation, likelihood, and knot placement}
\label{subsec:spline_reconstruction}

We parameterize the distance--redshift relation by the values of $\log r_p$ at $K$ redshift knots $\{z_0 = 0,\, z_1,\, \ldots,\, z_{K-1}\}$. The first knot is fixed kinematically: $y_0 \equiv \log r_p(z_0=0) = 0$, since $r(z) \to cz/H_0$ as $z \to 0$ in any FLRW cosmology. The remaining $K-1$ knot values $\bm{y} = (y_1, \ldots, y_{K-1})$ are the free parameters constrained by the data, and together with their Gaussian covariance form the compressed data product used for downstream cosmological analyses. At any redshift $z$, $\log r_p(z)$ is obtained by a cubic spline interpolation through the knots, which is linear in the knot values:
\begin{equation}
    \log r_p(z) = \sum_{k=0}^{K-1} S_k(z)\, y_k,
    \label{eq:spline_basis}
\end{equation}
where $S_k(z)$ are the cubic-spline basis functions, depending only on the knot positions $\{z_k\}$ and the natural boundary conditions ($y'' = 0$ at $z_0$ and $z_{K-1}$).

The theoretical distance modulus is given by $\mu_{\rm th}(z) = 5\log_{10}(d_L(z)/\mathrm{Mpc}) + 25$, with $d_L(z) = (1+z)\,r(z)$ in a flat universe. Substituting our $r_p$ parameterization (Eq.~\ref{eq:cadillac_transform}) gives
\begin{equation}
    \mu_{\rm th}(z) = 5\log_{10} \left [(1+z)^{1-{\mathcal{W}(z)}}\,z\,r_p(z) \cdot \frac{c H_0^{-1}}{\mathrm{Mpc}}\right] + 25,
    \label{eq:mu_th_rp}
\end{equation}
Combining Eqs.~(\ref{eq:spline_basis}) and (\ref{eq:mu_th_rp}), the per-SN distance modulus is exactly linear in $\bm{y}$:
\begin{equation}
    \mu_{\rm th}(z) =  \mathcal{M} + \nu(z) + \frac{5}{\ln 10}\sum_{k=0}^{K-1} S_k(z)\, y_k,
    \label{eq:mu_linear_in_y}
\end{equation}
where the global nuisance offset $\mathcal{M}$ absorbing the $H_0$ dependence and the cosmology-free function $\nu(z)$ are
\begin{equation}
\begin{aligned}
    \mathcal{M} &\equiv 5\log_{10}\!\left(\frac{cH_0^{-1}}{\mathrm{Mpc}}\right) + 25, \\
    \nu(z) &\equiv 5\log_{10}(z) + 5\bigl(1 - \mathcal{W}(z)\bigr)\log_{10}(1+z).
\end{aligned}
\label{eq:M_nu_defs}
\end{equation}
Both $\mu_{\rm th}(z)$ and $\nu(z)$ diverge as $5\log_{10}(z) \to -\infty$ at $z \to 0$. The difference is finite and proportional to $\log r_p(z)$, consistent with the FLRW boundary condition $\log r_p(0) = 0$ that sets $y_0$. In practice $\nu(z)$ is only evaluated at the SN redshifts $\{z_i\}$, which all satisfy $z_i > 0$, so this divergence never enters the numerical pipeline.

We analytically marginalize over $\mathcal{M}$ and obtain the following modified $\chi^2$ to be minimized,
\begin{equation}
    \chi^2_{\rm marg} = (\Delta \boldsymbol{\mu})^T \mathbf{C}^{-1} (\Delta \boldsymbol{\mu}) - \frac{\left[ (\Delta \boldsymbol{\mu})^T \mathbf{C}^{-1} \mathbf{1} \right]^2}{\mathbf{1}^T \mathbf{C}^{-1} \mathbf{1}},
    \label{eq:chi2_marg}
\end{equation}
with $\Delta \boldsymbol{\mu}$ the data residual against the $\mathcal{M}$-independent part of $\boldsymbol{\mu}_{\rm th}$ (Eqs.~\ref{eq:mu_linear_in_y} and \ref{eq:M_nu_defs}), and $\mathbf{C}$ is the full covariance matrix of $\boldsymbol{\mu}_{\rm obs}$. For the flux-averaged analysis, individual SNe are compressed into redshift bins following Paper~I \cite{PaperI}. The bin-level likelihood, including the model-dependent $\ln \det \bar{\mathbf{C}}$ term from the parameter dependence of the projection matrix, is treated in Sec.~\ref{subsec:fa_newton}.

The $\mathcal{M}$ marginalization makes explicit that the data constrain only the \emph{shape} of $\log r_p(z)$: a uniform shift $y_k \to y_k + \delta$ at all knots is exactly compensated by $\mathcal{M} \to \mathcal{M} - 5\delta\log_{10} e$, so the overall amplitude of $\{y_k\}$ is degenerate with the nuisance parameter and does not enter the cosmological inference. 

We use 12 knots indexed $z_0, z_1, \ldots, z_{11}$ per dataset. Pantheon, Pantheon+, and Union3 ($z_{\rm max} \approx 2.26$) share the same grid; DES-Dovekie has a lower redshift maximum ($z_{\rm max} \approx 1.14$), so its top three knots are shifted accordingly. The two grids are listed in Table~\ref{tab:knots}. Each knot interval contains $\geq 15$--$20$ SNe.

\begin{table}[h]
\centering
\caption{Spline knot redshifts $z_k$ for the four SNe~Ia compilations. Pantheon, Pantheon+, and Union3 share the same 12-knot grid covering $0 \leq z \leq 2.27$; DES-Dovekie uses a 12-knot grid truncated at $z_{11} = 1.15$ to match its $z_{\rm max} \approx 1.14$. Only 11 knots with $z_k > 0$ carry cosmological information and are used for downstream cosmological inference.}
\label{tab:knots}
\renewcommand{\arraystretch}{1.2}
\setlength{\tabcolsep}{4pt}
\small
\begin{tabular}{lcccccccccccc}
\toprule
$k$ & 0 & 1 & 2 & 3 & 4 & 5 & 6 & 7 & 8 & 9 & 10 & 11 \\
\midrule
Pantheon  & \multirow{3}{*}{0} & \multirow{3}{*}{0.05} & \multirow{3}{*}{0.15} & \multirow{3}{*}{0.30} & \multirow{3}{*}{0.40} & \multirow{3}{*}{0.50} & \multirow{3}{*}{0.60} & \multirow{3}{*}{0.70} & \multirow{3}{*}{0.80} & \multirow{3}{*}{1.00} & \multirow{3}{*}{1.60} & \multirow{3}{*}{2.27} \\
Pantheon+ &  &  &  &  &  &  &  &  &  &  &  &  \\
Union3    &  &  &  &  &  &  &  &  &  &  &  &  \\
\midrule
DES-Dovekie & 0 & 0.05 & 0.15 & 0.30 & 0.40 & 0.50 & 0.60 & 0.70 & 0.80 & 0.90 & 1.00 & 1.15 \\
\bottomrule
\end{tabular}
\end{table}

\subsection{Without flux averaging: closed-form estimator}
\label{subsec:closed_form}

The spline forward model in Eq.~(\ref{eq:mu_linear_in_y}) is exactly linear in the knot values $\bm{y}$: $\boldsymbol{\mu}_{\rm th} = \boldsymbol{\nu} + \mathcal{M}\,\mathbf{1} + \mathbf{A}\,\bm{y}$, where $\boldsymbol{\nu}$ is the cosmology-independent function (Eq.~\ref{eq:M_nu_defs}) collected over the SN redshifts and $\mathbf{A}$ is the base matrix of cubic-spline basis functions evaluated at those redshifts. With the standard SN~Ia assumption that $\boldsymbol{\mu}_{\rm obs}$ is Gaussian with covariance $\mathbf{C}_\mu$, this is a {\it linear-Gaussian model}: after analytic $\mathcal{M}$ marginalization (Eq.~\ref{eq:chi2_marg}), the posterior on $\bm{y}$ is exactly Gaussian \cite{NumericalRecipes}, with mean and covariance given by the best linear unbiased estimator (BLUE),
\begin{equation}
    \hat{\bm{y}} = (\mathbf{A}^T \mathbf{M} \mathbf{A})^{-1}\, \mathbf{A}^T \mathbf{M}\, (\bm{\mu}_{\rm obs} - \bm{\nu}),
    \qquad
    \boldsymbol{\Sigma}_y = (\mathbf{A}^T \mathbf{M} \mathbf{A})^{-1}.
    \label{eq:closed_form_main}
\end{equation}
The base matrix $\mathbf{A}$ and the $\mathcal{M}$-marginalization projection $\mathbf{M}$ are
\begin{align}
    A_{ik} &= \frac{5}{\ln 10}\,S_k(z_i), \quad i = 1,\ldots,N_{\rm SN}, \ k = 1,\ldots,K-1, \label{eq:design_matrix_A}\\
    \mathbf{M} &= \mathbf{C}_\mu^{-1} - \frac{\mathbf{C}_\mu^{-1}\mathbf{1}\mathbf{1}^T \mathbf{C}_\mu^{-1}}{\mathbf{1}^T \mathbf{C}_\mu^{-1}\mathbf{1}}. \label{eq:M_projection}
\end{align}
The $k=0$ column is dropped because $y_0 = 0$ is fixed kinematically. In practice $\mathbf{A}$ is built numerically column by column: for each $k$, set $\bm{y} = \mathbf{e}_k$ (the indicator vector, with $y_0 = 0$ held fixed), pass the result through any natural cubic-spline solver (e.g., \texttt{DataInterpolations.jl}), and evaluate at the SN redshifts $\{z_i\}$. No closed-form expression for $S_k(z)$ is needed; the spline solver does one tridiagonal solve per column in $\mathcal{O}(N)$ operations.

Operationally, no iteration or MCMC is needed: build $\mathbf{A}$ from the spline basis at the SN redshifts, evaluate $\boldsymbol{\nu}$ from $\{z_i\}$, compute $\mathbf{M}$ from $\mathbf{C}_\mu$, and solve Eq.~(\ref{eq:closed_form_main}). The total cost is $\mathcal{O}(10^{-2})$\,s per dataset, four-to-five orders of magnitude faster than the equivalent MCMC chain, and the BLUE agrees with direct NUTS sampling to within Monte-Carlo noise (Sec.~\ref{subsec:validation}).

\subsection{With flux averaging: iterative Newton estimator}
\label{subsec:fa_newton}
At $z \gtrsim 1$ weak gravitational lensing produces a non-Gaussian magnification PDF in magnitude space, biasing magnitude-averaged distances through Jensen's inequality. Averaging in linear flux space removes the bias \cite{Wang2000, Wang1999}. Following \cite{ZhaiWang2019, WangMukherjee2004} and Paper~I~\cite{PaperI}, the flux-averaging procedure at a fixed cosmology $\bm{y}$ has four steps: (i) convert each SN magnitude to flux $F_i = 10^{-0.4(\mu_{{\rm obs},i} - 25)}$; (ii) form the model-corrected scaled luminosity $\mathcal{L}_i = F_i\, d_L^2(z_i; \bm{y})$; (iii) bin SNe in redshift and average within bin $b$, $\bar{\mathcal{L}}_b = N_b^{-1}\sum_{i\in b}\mathcal{L}_i$ at $\bar z_b = N_b^{-1}\sum_{i\in b} z_i$; (iv) convert back to a flux-averaged distance modulus
\begin{equation}
    \bar\mu_b = 25 - 2.5\log_{10}\!\left[\bar{\mathcal{L}}_b / d_L^2(\bar z_b; \bm{y})\right],
    \label{eq:wang_fa_mu}
\end{equation}
with bin covariance $\bar{\mathbf{C}}(\bm{y}) = \mathbf{P}\,\mathbf{C}_\mu\,\mathbf{P}^T$, $P_{bi} = \mathcal{L}_i/(N_b \bar{\mathcal{L}}_b)$. We use a hybrid bin layout: $\sim 40$ equal-count bins below $z = 0.8$, plus three uniform bins per knot interval above $z = 0.8$. This places three uniformly-spaced flux-averaged bins within each knot interval above $z = 0.8$ (so bin widths differ between intervals, matching the varying knot spacing), so the cubic-spline fit at every high-$z$ knot has local data support on both sides despite the low SN density at high redshift. The binning choice differs from the one used in Paper~I \citep{PaperI}, but is applied uniformly to every flux-averaged analysis in this work.

Both $\bar{\boldsymbol{\mu}}(\bm{y})$ and $\bar{\mathbf{C}}(\bm{y})$ depend on $\bm{y}$, so the model is no longer linear in $\bm{y}$ and the BLUE of Sec.~\ref{subsec:closed_form} no longer applies in one solve. The bin-level negative log-likelihood is
\begin{equation}
    -2\ln\mathcal{L}(\bm{y}) = \chi^2_{\rm bin}(\bm{y}) + \ln\det \bar{\mathbf{C}}(\bm{y}) + \mathrm{const},
    \label{eq:logL_FA}
\end{equation}
with $\chi^2_{\rm bin}(\bm{y})$ the BLUE-style quadratic of \refeq{chi2_marg} evaluated on the bin-level data $(\bar{\boldsymbol{\mu}}, \bar{\mathbf{C}})$. The extra $\ln\det \bar{\mathbf{C}}(\bm{y})$ term is the model-dependent normalisation of the binned Gaussian: it is absent in Sec.~\ref{subsec:closed_form} because the per-SN covariance $\mathbf{C}_\mu$ is parameter-independent, but here $\bar{\mathbf{C}}$ varies with $\bm{y}$ through the flux-averaging projection $\mathbf{P}(\bm{y})$. We minimize $-2\ln\mathcal{L}$ by Newton iteration,
\begin{equation}
    \bm{y}^{(j+1)} = \bm{y}^{(j)} - \mathbf{H}^{-1}(\bm{y}^{(j)})\,\nabla(-2\ln\mathcal{L})(\bm{y}^{(j)}),
    \label{eq:newton_FA}
\end{equation}
where the Hessian
\begin{equation}
    \mathbf{H}(\bm{y}) \equiv \nabla\nabla^{T}\!\bigl[-2\ln\mathcal{L}(\bm{y})\bigr]
    \label{eq:hessian_FA}
\end{equation}
and the gradient $\nabla(-2\ln\mathcal{L})$ are computed at each step by automatic differentiation through the flux-averaging pipeline. The iteration converges in two Newton steps to $\max|\Delta y| \le 10^{-7}$ on Pantheon, with posterior covariance ${\boldsymbol{\Sigma}_y} = \tfrac{1}{2}\mathbf{H}^{-1}$. The total cost is $\mathcal{O}(10)$\,s per dataset, compared to a few hours for direct NUTS sampling of the full nonlinear flux-averaged likelihood (Sec.~\ref{subsec:validation}).

\subsection{Practical recipe}
\label{subsec:recipe}

The compressed $\log r_p$ is designed to simple and straightforward to use.
Recipe~A extracts the Gaussian posterior $(\hat{\bm{y}}, \boldsymbol{\Sigma}_y)$ from a SNe~Ia compilation; Recipe~B uses it in a downstream cosmological MCMC. Together they make explicit which nuisance parameters are analytically marginalized and which physical parameters are sampled.

\subsubsection{Recipe~A: compressing SN distance moduli into $\log r_p$ knots}

\textit{Inputs:} Per-SN distance moduli $\{\mu_{{\rm obs},i}\}_{i=1}^{N_{\rm SN}}$ with the full $N_{\rm SN}\times N_{\rm SN}$ covariance $\mathbf{C}_\mu$, the redshift-knot grid $\{z_k\}_{k=0}^{11}$ from Table~\ref{tab:knots}, and the rescaling $\mathcal{W}(z)$ in Eq.~(\ref{eq:cadillac_transform}).

\noindent
\textit{Steps:}
\begin{enumerate}
    \item Build the $N_{\rm SN}\times 11$ base matrix $\mathbf{A}$ from Eq.~(\ref{eq:design_matrix_A}), with $S_k(z)$ the natural cubic-spline basis on the 11 free knots.
    \item Build the cosmographic baseline $\boldsymbol{\nu}$ by evaluating $\nu(z_i)$ of Eq.~(\ref{eq:M_nu_defs}) at each SN redshift.
    \item Build the $\mathcal{M}$-marginalization projection $\mathbf{M}$ of Eq.~(\ref{eq:M_projection}); this step analytically marginalizes the nuisance parameter combining $H_0$ and $M_B$.
    \item Solve the closed-form BLUE of Eq.~(\ref{eq:closed_form_main}) for the mean $\hat{\bm{y}}$ and covariance $\boldsymbol{\Sigma}_y$. For the flux-averaged variant, replace this step with the iterative Newton procedure of Sec.~\ref{subsec:fa_newton} on the bin-level data; convergence takes two Newton steps.
\end{enumerate}

\noindent
\textit{Output:} $(\hat{\bm{y}}, \boldsymbol{\Sigma}_y)$, the Gaussian posterior on the 11 free $\log r_p$ knots, with $\log r_p(z=0) = 0$ fixed by FLRW kinematics. No MCMC is required and no cosmological parameters are sampled.

\subsubsection{Recipe~B: cosmological inference with $\log r_p$ and covariance}

\textit{Inputs:} $(\hat{\bm{y}}, \boldsymbol{\Sigma}_y)$ from Recipe~A or directly from Appendix~\ref{app:compressed_data}, and the set of cosmological model parameters $\bm{\theta}$ through the expansion rate $E(z;\bm{\theta}) \equiv H(z;\bm{\theta})/H_0$.

\noindent
\textit{Steps.} At each step of the cosmological MCMC:
\begin{enumerate}
    \item Compute the theory $\log r_p$ at the 11 redshift knots $z_k$, using the same $\mathcal{W}(z)$ as Recipe~A:
    \begin{equation*}
        y_k^{\rm th}(\bm{\theta}) = \log\!\left[\frac{1}{z_k}\int_0^{z_k}\!\frac{dz'}{E(z';\bm{\theta})}\right] + \mathcal{W}(z_k)\,\log(1+z_k).
    \end{equation*}
    The $H_0$ and $M_B$ dependence are absorbed into the kinematic anchor $y_0 = 0$ and the $\mathcal{M}$ marginalization of Recipe~A, respectively, so $\bm{y}^{\rm th}(\bm{\theta})$ depends only on $\bm{\theta}$.
    \item Evaluate the SNe~Ia log-likelihood,
    \begin{equation*}
        -2\ln\mathcal{L}_{\rm SN}(\bm{\theta}) = (\hat{\bm{y}} - \bm{y}^{\rm th}(\bm{\theta}))^T\,\boldsymbol{\Sigma}_y^{-1}\,(\hat{\bm{y}} - \bm{y}^{\rm th}(\bm{\theta})).
    \end{equation*}
    \item Sample $\bm{\theta}$. For joint analyses with CMB or BAO, additional parameters such as $H_0$ and $\omega_b$ enter through their own likelihoods; the SNe term remains a pure Gaussian on $\bm{\theta}$.
\end{enumerate}

\noindent
\textit{Output:} Posterior samples of $\bm{\theta}$. A single MCMC step requires one 11-dimensional matrix-vector product and one one-dimensional redshift integral per knot, independent of $N_{\rm SN}$.

The three flat-universe cosmological models we apply Recipe~B to, following Paper~I~\cite{PaperI}, have $E(z;\bm{\theta})$ as
\begin{equation}
E(z) =
\begin{cases}
\left[ \Omega_{m}(1+z)^3 + (1-\Omega_m) \right]^{1/2}, & \text{flat } \Lambda\text{CDM} \\[4pt]
\left[ \Omega_{m}(1+z)^3 + (1-\Omega_m)\,(1+z)^{3(1+w_0+w_a)} \exp\!\left(-\frac{3 w_a z}{1+z}\right) \right]^{1/2}, & \text{flat } w_0 w_a\text{CDM} \\[4pt]
\left[ \Omega_{m}(1+z)^3 + (1-\Omega_m)\,X(z)\right]^{1/2}, & X(z) \text{ dark energy}
\end{cases}
\label{eq:Ez_models}
\end{equation}
For $w_0 w_a$CDM~\cite{Chevallier2001, Linder2003} the dark-energy equation of state is $w(a) = w_0 + w_a(1-a)$, giving $\bm{\theta} = (\Omega_m, w_0, w_a)$. For the non-parametric model $X(z) \equiv \rho_{\rm DE}(z)/\rho_{\rm DE}(0)$ is sampled at five free knots $z \in \{1/3, 2/3, 1, 4/3, 2.33\}$ with $X(0) = 1$ fixed and intermediate values from linear interpolation in $z$, giving $\bm{\theta} = (\Omega_m, \omega_b, H_0, X_1, \ldots, X_5)$. SNe~Ia alone do not place meaningful constraints on the latter two models, so for those we combine the SNe likelihood of Recipe~B with the Planck~2015 CMB distance priors of \cite{WangDai2016} and the DESI~DR2 BAO measurements~\cite{DESIDR2}. The full priors and sampling settings follow Paper~I~\cite{PaperI}. The losslessness of the two-step pipeline (Recipe~A then Recipe~B) against the per-SN reference is demonstrated across all datasets and flux-averaging configurations and these three cosmological models in Sec.~\ref{subsec:cosmo_from_rp}.

\section{Results}
\label{sec:results}
Following the recipe in Sec.~\ref{subsec:recipe}, we present the compression of per-SN distance moduli $\bm\mu$ into $\log r_p$ values at 11 redshift knots for the same four SNe~Ia compilations as in Paper~I~\cite{PaperI}: Pantheon~\cite{Scolnic2018}, Pantheon+~\cite{Scolnic2022}, DES-Dovekie~\cite{DESY5_2024, Dovekie2025}, and Union3~\cite{Rubin2023}. We then perform the cosmological analysis based on these 11 $\log r_p$ knots and their covariance matrix. For the flux-averaged analyses, we apply a quality cut $\sigma_\mu < 0.5$ to retain 1726 SNe of DES-Dovekie. Union3 is pre-binned at the catalog level and flux-averaging cannot be applied. We refer to Paper~I~\cite{PaperI} for further details on data preparation and quality cuts.

In Sec.~\ref{subsec:validation} we validate our estimators for $\log r_p$ and covariance matrix by comparing against the direct MCMC sampling in both flux-averaged and not-flux-averaged cases. We also show that $\log r_p$ posterior is Gaussian while the $r_p$ posterior is not at the highest redshifts where the uncertainties are the largest. We then present the compressed $\log r_p$ results for all four datasets in Sec.~\ref{subsec:rp_reconstructions}. Finally, in Sec.~\ref{subsec:cosmo_from_rp} we provide an apples-to-apples comparison between the cosmological inference using $\log r_p$ and using $\mu(z)$, showing that the $\log r_p$ pipeline can accurately reproduce the cosmological parameter posteriors of Paper~I but with much lower computational cost. The compressed $\{\log r_p(z_k)\}$ values and their covariance matrices are tabulated in Appendix~\ref{app:compressed_data} to enable downstream cosmological analyses using the methodology presented in this paper.

\subsection{Validation of estimators and Gaussian posterior}
\label{subsec:validation}

\begin{figure*}[h]
    \centering
    \includegraphics[width=\linewidth]{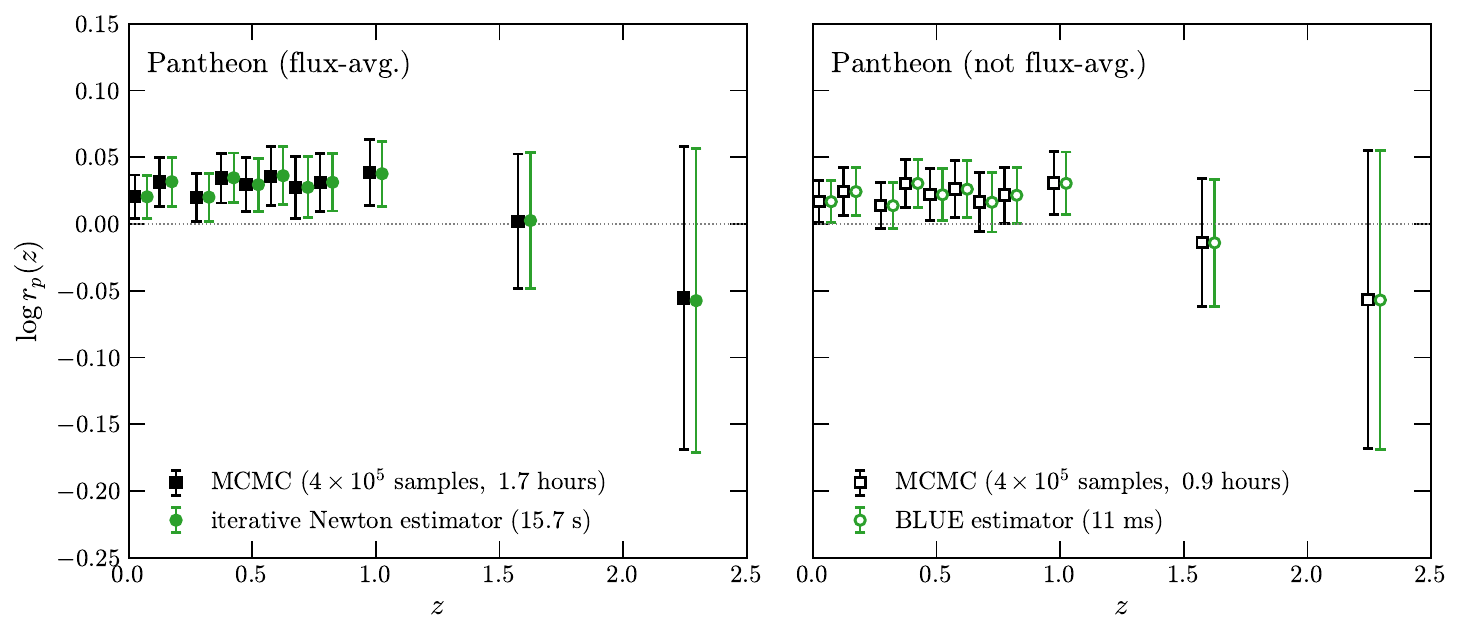}
    \caption{Validation of the analytic $\log r_p$ estimators against MCMC, for Pantheon. {\it Left}: flux-averaged data; the iterative Newton estimator (green circles, $16$\,s) overlays the full nonlinear flux-averaged NUTS posterior ($4\times 10^5$ samples, $\sim$1.7\,h, black squares). {\it Right}: not-flux-averaged data; the closed-form BLUE estimator (open green circles, $11$\,ms) overlays the direct NUTS posterior (open black squares, $\sim$0.9\,h).
    The two sets of measurements overlap completely, and are shifted by $\pm 0.025$ in $z$ for visualization.}
    \label{fig:logrp_validation_2panel}
\end{figure*}

We first verify, on Pantheon, that our two estimators (closed-form BLUE for the not-flux-averaged data, and iterative Newton for the flux-averaged data) reproduce the full Bayesian posterior obtained by direct MCMC sampling of the eleven $\log r_p$ knot values. The MCMC takes four chains of $10^5$ NUTS samples and runs for $\sim 0.9$\,h (without flux averaging) and $\sim 1.7$\,h (with flux averaging). The corresponding estimators complete in $11$\,ms and $16$\,s respectively.

Figure~\ref{fig:logrp_validation_2panel} compares the two estimators against MCMC at all eleven knots. The agreement is well within the chains' Monte-Carlo sampling noise per knot in both panels. In the not-flux-averaged case this is the linear-Gaussian theorem made explicit: the reconstruction of $\log r_p$ through cubic spline interpolation is linear in distance modulus $\mu$ and the SNe~Ia likelihood is Gaussian in $\bm{\mu}_{\rm obs}$, so the posterior mean and the covariance matrix obtained from MCMC is guaranteed to converge to the BLUE result (Sec.~\ref{subsec:closed_form}). The flux-averaging is intrinsically nonlinear in $\mu$ so there is no closed-form estimator, but the iterative Newton procedure converges to the full MCMC results within sampling noise, in which flux averaging is dynamically applied at each step. The analytic estimators thus provide a posterior numerically consistent with MCMC at four to five orders of magnitude lower cost.

\begin{figure}[t]
    \centering
    \includegraphics[width=0.9\linewidth]{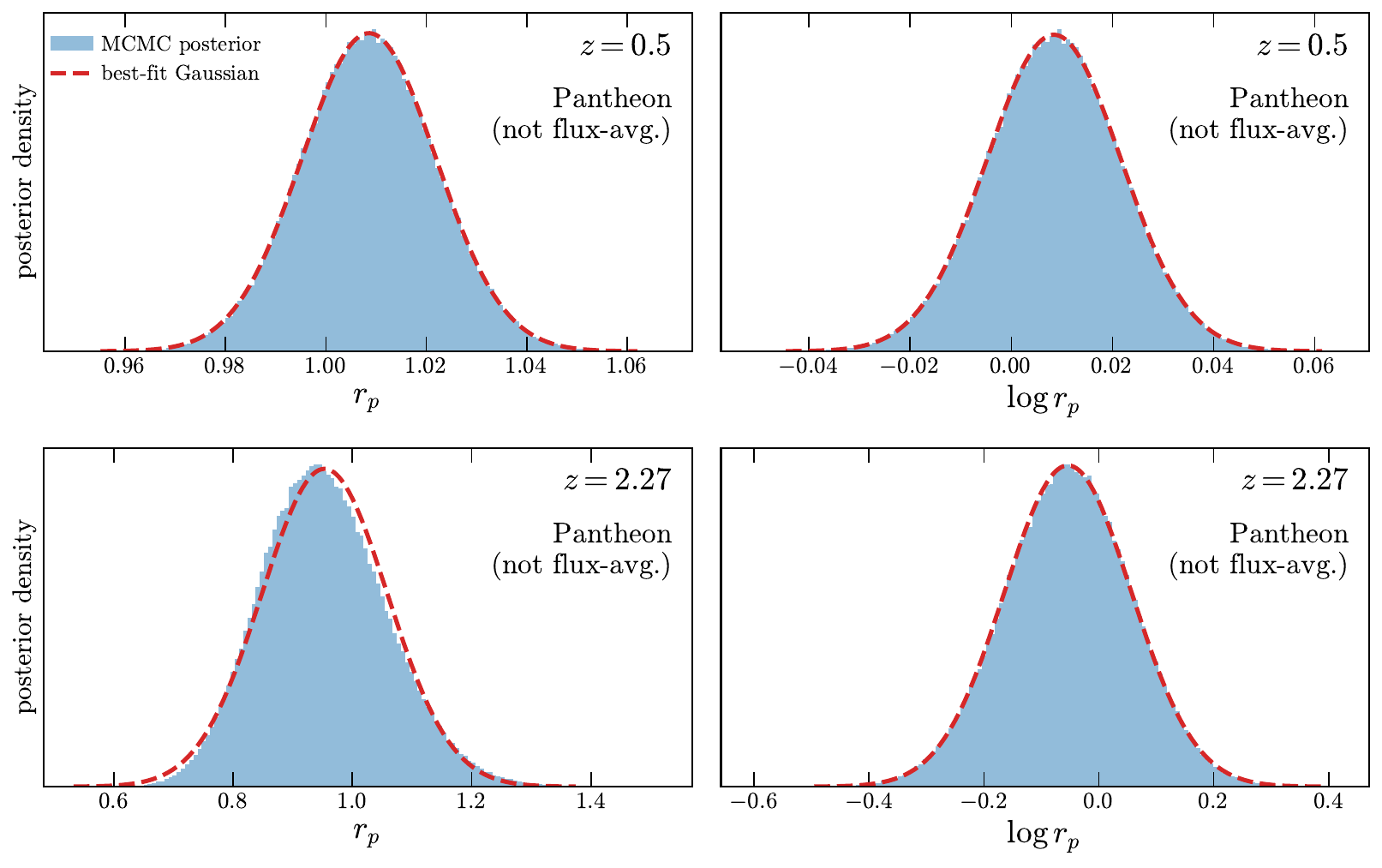}
    \caption{Marginal posteriors of $r_p$ (left column) and $\log r_p$ (right column) at two redshift knots, $z = 0.5$ (top row) and $z = 2.27$ (bottom row), for the Pantheon without flux averaging. Blue histograms represent the MCMC posteriors. Red dashed curves are the best-fit Gaussian to each histogram. At $z = 0.5$ the per-knot uncertainty is small enough that the $r_p$ and $\log r_p$ marginals are visually indistinguishable. At $z = 2.27$ the per-knot uncertainty is large, and the $r_p$ marginal develops a clear positive skew while the $\log r_p$ marginal stays Gaussian.}
    \label{fig:gaussianity_panel}
\end{figure}

The check above establishes mean--and--variance agreement between the estimators and MCMC. We also test the Gaussianity of the $\log r_p$ posterior, which justifies treating it as a Gaussian likelihood in the downstream cosmology analysis. Sec.~\ref{subsec:closed_form} predicts that the marginal posterior on the $\log r_p$ knots is exactly Gaussian under the standard SN~Ia likelihood, which implies that the posterior of $r_p = e^{\log r_p}$ is log-normal, with skewness that grows with the per-knot uncertainty and is therefore most visible at the highest-redshift knot, where the data constrain $r_p$ least. We verify this directly in Fig.~\ref{fig:gaussianity_panel}, which compares the marginal posterior of $r_p(z_k)$ and $\log r_p(z_k)$ against a best-fit Gaussian at two representative knots ($z = 0.5$ and $z = 2.27$) of the Pantheon SN Ia data set. At low redshift the per-knot uncertainty is small enough that the two parameterizations are indistinguishable. At the highest knot, the $r_p$ posterior develops a clear positive skew while the $\log r_p$ marginal at the same knot remains Gaussian.

Working in $\log r_p$ rather than $r_p$ is therefore the optimal choice for accuracy and consistency for downstream cosmological analysis. Propagating an $r_p$-knot mean and covariance as if it were a Gaussian likelihood would misrepresent the data at the high-$z$ knots where the per-knot uncertainty is large. The $\log r_p$ marginal is exactly Gaussian by construction (Sec.~\ref{subsec:closed_form}), so the downstream Gaussian-likelihood treatment of $\{\log r_p(z_k)\}$ and their covariance matrix is completely consistent with the standard distance-modulus-based analysis.

The Gaussianity of $\log r_p$ rests on an assumption \emph{plus} an additional statistical robustness advantage. The standard SN~Ia likelihood assumes $\boldsymbol{\mu}_{\rm obs}$ is Gaussian, and under that assumption the linear-Gaussian theorem (Sec.~\ref{subsec:closed_form}) makes each $\log r_p$ knot \emph{exactly} Gaussian. On the other hand, the actual per-SN distribution of $\mu_{\rm obs}$ is itself only assumed Gaussian and not directly verifiable in the data. The linear compression step nevertheless provides a central-limit-theorem guarantee that the $\log r_p$ knots are closer to Gaussian than the input $\boldsymbol{\mu}_{\rm obs}$. Each knot is a weighted linear combination of $\mathcal{O}(10^2)$ per-SN distance moduli , so any departure of the underlying per-SN error distribution from the assumed Gaussian form is suppressed in $\hat y_k$ by $\sim 1/\sqrt{N}$. In this sense the $\log r_p$ representation is the more robust statistically: the Gaussian-likelihood treatment we adopt in the downstream cosmological inference is consequently a better-justified approximation for $\log r_p$ than it is for $\boldsymbol{\mu}_{\rm obs}$.

\subsection{Compression of distance measurement}
\label{subsec:rp_reconstructions}

\begin{figure*}
    \centering
    \includegraphics[width=0.9\linewidth]{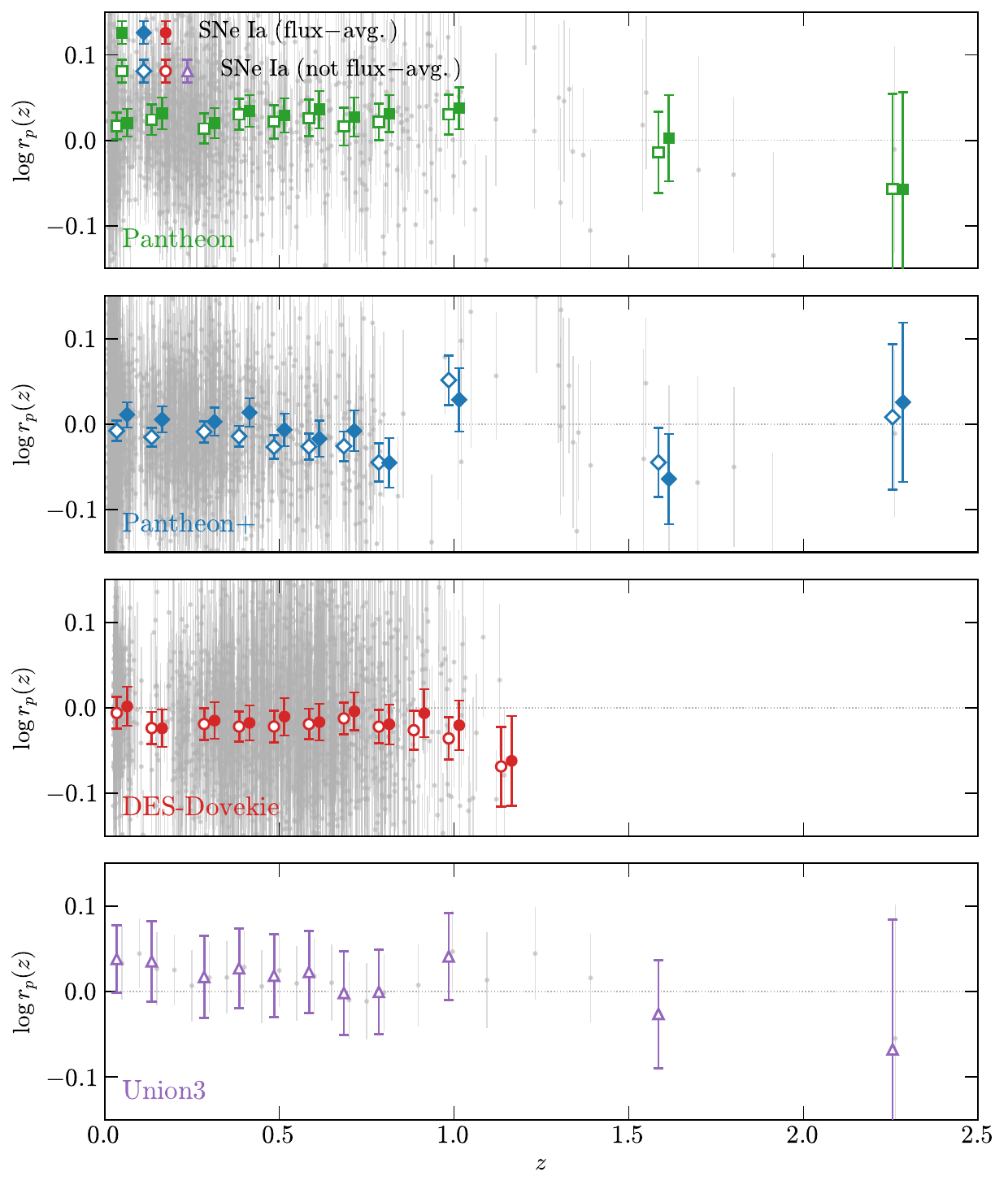}
    \caption{Closed-form $\log r_p(z)$ reconstruction from the four SNe~Ia compilations. Open markers show the not-flux-averaged BLUE knots, and filled markers show the iterative-Newton flux-averaged knots (Union3 is pre-binned and has no flux-avg.\ mode). The two sets are shifted by $\pm 0.015$ in $z$ for visualization only. The faint grey error bars in the background are the individual SN distance moduli rescaled to $\log r_p$, which the colored markers compress into eleven knot values. The knot values are tabulated for each dataset in Appendix~\ref{app:compressed_data}.}
    \label{fig:rp_flatLCDM}
\end{figure*}

Figure~\ref{fig:rp_flatLCDM} presents the $\log r_p(z)$ results for all four datasets in both not-flux-averaged and flux-averaged configurations using the estimators of Sec.~\ref{subsec:closed_form} and Sec.~\ref{subsec:fa_newton}. The different absolute $\log r_p$ levels visible across the four panels reflect each dataset's preferred $\hat{\mathcal{M}}$, the global offset already analytically marginalized at the compression step (Sec.~\ref{subsec:closed_form}). Since $\mathcal{M}$ is degenerate with $H_0$ and $M_B$, these per-dataset offsets have no effect on the cosmological parameter posteriors. The full uncertainty on the 11 knot values, including the amplitude information contributed by SNe in the low-$z$ anchor region $[z_0, z_1] = [0, 0.05]$ via the FLRW kinematic constraint $\log r_p(0) = 0$, is captured by the $11 \times 11$ covariance matrix $\boldsymbol{\Sigma}_y$, which the downstream Gaussian likelihood weights correctly through $\boldsymbol{\Sigma}_y^{-1}$.

We also plot the original $\boldsymbol{\mu}_{\rm obs}$ data and their uncertainties as the faint grey error bars, rescaled to the same $\log r_p$ axis. The per-SN scatter is large, particularly at high redshift, and the underlying distance--redshift trend is essentially invisible from the grey cloud alone. The eleven coloured knot values, each effectively averaging $\mathcal{O}(10^2)$ SNe through the BLUE projection, compress this cloud into a clean reconstruction that makes the data-driven shape of $\log r_p(z)$ directly readable, with per-knot uncertainties roughly an order of magnitude tighter than per individual SN.

The largest per-knot uncertainties appear at the highest-$z$ knots ($z \gtrsim 0.8$), where the flux-averaging weighting is most sensitive to the SNe~Ia with the largest scatter.


\subsection{Cosmological inference with compressed data}
\label{subsec:cosmo_from_rp}
Sec.~\ref{subsec:validation} establishes that the estimators and the direct MCMC of $\log r_p$ coincide at the per-knot level, and that the $\log r_p$ posterior is genuinely Gaussian. The remaining question---and the central claim of this paper---is whether the resulting eleven Gaussian-distributed $\log r_p$ values together carry the same cosmological-parameter information that the original per-SN analysis would have produced. We test this by feeding the compressed product $(\hat{\mathbf y}, \boldsymbol\Sigma_y)$ in Appendix~\ref{app:compressed_data} into a downstream MCMC for the same three cosmological models analysed in Paper~I~\cite{PaperI} (flat $\Lambda$CDM, flat $w_0 w_a$CDM, and the non-parametric dark-energy density $X(z) \equiv \rho_{\rm DE}(z)/\rho_{\rm DE}(0)$), and comparing every cosmological parameter posterior against the corresponding per-SN reference. The downstream cosmology in this section is therefore a validation test of the compression rather than a fresh constraint, but the dramatic speed-up the compression unlocks makes the same chains practical to rerun for any future dark-energy model.

\subsubsection{Flat $\Lambda$CDM}
We start with the simplest case, flat $\Lambda$CDM. We sample $\Omega_m$ alone (SNe-only), running $10^6 \times 4$ NUTS chains, which complete in approximately one minute per dataset. Results are summarized in Table~\ref{tab:Om_from_rp}.

\begin{table*}[t]
\centering
\caption{Flat $\Lambda$CDM $\Omega_m$ from the closed-form $\log r_p(z)$ posterior with downstream MCMC ($10^6 \times 4$ samples), compared with the full distance--modulus fits ($10^5 \times 4$ samples). The not-flux-averaged $\mu(z)$ entries are the original Paper~I~\cite{PaperI} chains; the flux-averaged $\mu(z)$ entries for Pantheon+ and DES-Dovekie are recomputed in this work using the hybrid binning choice of Sec.~\ref{subsec:fa_newton} so that they share the same flux-averaging configuration as the $\log r_p$-side analysis. Each $\log r_p$ value is followed in parentheses by the fractional difference $\Delta\Omega_m/\Omega_m \equiv \left(\Omega_m^{\log r_p} - \Omega_m^{\mu(z)}\right)/\Omega_m^{\mu(z)}$.}
\label{tab:Om_from_rp}
\setlength{\tabcolsep}{4pt}
\small
\begin{tabular}{@{}lcccc@{}}
\toprule
 & \multicolumn{2}{c}{$\log r_p$ (this work) ($\Delta\Omega_m/\Omega_m$)} & \multicolumn{2}{c}{$\mu(z)$~\citep{PaperI}} \\
\cmidrule(lr){2-3} \cmidrule(lr){4-5}
Dataset & not flux-avg. & flux-avg. & not flux-avg. & flux-avg. \\
\midrule \midrule
Pantheon      & $0.301 \pm 0.022$ ($-0.1\%$) & $0.292 \pm 0.022$ ($-0.1\%$) & $0.301 \pm 0.022$ & $0.292 \pm 0.023$ \\
\midrule
Pantheon+     & $0.331 \pm 0.018$ ($-0.1\%$) & $0.336 \pm 0.028$ ($-0.9\%$) & $0.332 \pm 0.018$ & $0.339 \pm 0.029$ \\
\midrule
DES-Dovekie   & $0.329 \pm 0.015$ ($+0.1\%$) & $0.324 \pm 0.022$ ($-0.6\%$) & $0.329 \pm 0.015$ & $0.326 \pm 0.022$ \\
\midrule
Union3        & $0.358 \pm 0.027$ ($-0.04\%$) & ---  & $0.358 \pm 0.027$ & --- \\
\bottomrule
\end{tabular}
\end{table*}

\begin{figure}[t]
    \centering
    \includegraphics[width=\linewidth]{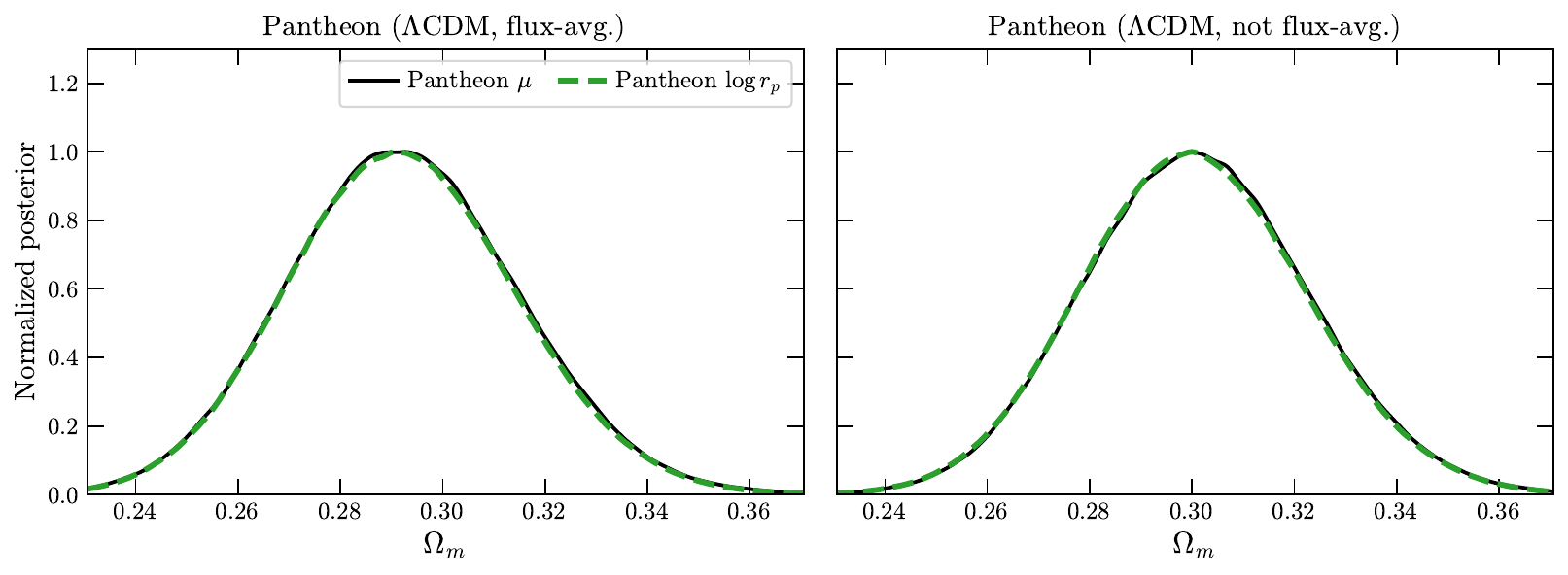}
    \caption{Posterior on $\Omega_m$ from the closed-form $\log r_p$ analysis (this work, green dashed) compared with the direct distance-modulus analysis of \citep{PaperI} (black solid) for Pantheon under flat $\Lambda$CDM. {\it Left:} not flux-averaged case. {\it Right:} flux-averaged case. The eleven $\log r_p$ data points reproduce the $1048$-SN $\mu$-based posterior precisely.}
    \label{fig:contour_Pantheon_LCDM}
\end{figure}

The $\log r_p$-derived $\Omega_m$ values reproduce the direct $\mu$-based fits to high precision. The not-flux-averaged chains agree to within $|\Delta\Omega_m|/\Omega_m \lesssim 0.1\%$ for every dataset. The flux-averaged chains shift slightly more: Pantheon+ with flux averaging differs from its $\mu$-based counterpart by $0.9\%$ and DES-Dovekie with flux averaging by $0.6\%$, while Pantheon with flux averaging stays at $0.1\%$. Figure~\ref{fig:contour_Pantheon_LCDM} compares the two $\Omega_m$ posteriors visually for Pantheon, where the eleven $\log r_p$ data points reproduce the full $1048$-SN $\mu$-based posterior to within sampling noise in both flux-averaged and not-flux-averaged modes.

\subsubsection{Flat $w_0w_a$CDM}
Once dark-energy freedom is included, SNe~Ia alone do not place meaningful constraints on cosmological parameters, so we combine the $\log r_p$ likelihood with the Planck~2015 CMB distance priors of \citep{WangDai2016} and DESI~DR2 BAO ($D_M/r_d$, $D_H/r_d$ at 12 redshift bins plus the BGS $D_V/r_d$ point) \citep{DESIDR2}. For the $w_0 w_a$CDM parameterization $w(a) = w_0 + w_a(1-a)$, we jointly sample $(\Omega_m, \omega_b, H_0, w_0, w_a)$. Chains comprise $2\times 10^5 \times 4$ NUTS samples (vs $10^5 \times 4$ in Paper~I); each runs in $\sim$10 minutes. Results are listed in Table~\ref{tab:cpl_from_rp}.

\begin{table*}[t]
\centering
\caption{Flat $w_0 w_a$CDM constraints from the closed-form $\log r_p(z)$ posterior plus the Planck~2015 CMB distance priors of \citep{WangDai2016} and DESI~DR2 BAO (top section), compared with the corresponding full distance--modulus chains (bottom section). The not-flux-averaged $\mu(z)$ chains are the original Paper~I~\cite{PaperI} runs; the flux-averaged $\mu(z)$ chains for Pantheon+ and DES-Dovekie are recomputed in this work using the hybrid binning choice of Sec.~\ref{subsec:fa_newton}, so the two sides differ only in whether the binned distance moduli are first compressed to $\log r_p$. In the top section each $\log r_p$ row is followed by a subrow giving the fractional difference 
$\Delta\theta/\theta\equiv \left(\theta^{\log r_p} - \theta^{\mu(z)}\right)/\theta^{\mu(z)}$. Each $\log r_p$-based chain uses $2\times 10^5 \times 4$ samples and completes in $\sim$10\,min per dataset.}
\label{tab:cpl_from_rp}
\setlength{\tabcolsep}{6pt}
\renewcommand{\arraystretch}{1.15}
\small
\begin{tabular}{@{}llccc@{}}
\toprule
\multicolumn{5}{c}{$\log r_p$ (this work)} \\
\midrule
Dataset & flux-avg. mode & $\Omega_m$ & $w_0$ & $w_a$ \\
\midrule \midrule
\multirow{4}{*}{\makecell[l]{Pantheon \\ {\scriptsize + Planck15 + DESI~DR2}}}
              & not flux-avg. & $0.304 \pm 0.007$ & $-0.925 \pm 0.070$ & $-0.34 \pm 0.25$ \\
              &               & $(-0.2\%)$ & $(-0.2\%)$ & $(+1.5\%)$ \\
              & flux-avg.     & $0.302 \pm 0.007$ & $-0.945 \pm 0.071$ & $-0.29 \pm 0.25$ \\
              &               & $(-0.1\%)$ & $(-0.1\%)$ & $(-0.2\%)$ \\
\midrule
\multirow{4}{*}{\makecell[l]{Pantheon+ \\ {\scriptsize + Planck15 + DESI~DR2}}}
              & not flux-avg. & $0.311 \pm 0.006$ & $-0.863 \pm 0.056$ & $-0.48 \pm 0.22$ \\
              &               & $(-0.1\%)$ & $(-0.04\%)$ & $(+0.1\%)$ \\
              & flux-avg.     & $0.311 \pm 0.008$ & $-0.874 \pm 0.076$ & $-0.43 \pm 0.26$ \\
              &               & $(-0.4\%)$ & $(-3.1\%)$ & $(+18.9\%)$ \\
\midrule
\multirow{4}{*}{\makecell[l]{DES-Dovekie \\ {\scriptsize + Planck15 + DESI~DR2}}}
              & not flux-avg. & $0.313 \pm 0.006$ & $-0.834 \pm 0.058$ & $-0.60 \pm 0.24$ \\
              &               & $(-0.1\%)$ & $(0.0\%)$ & $(-0.2\%)$ \\
              & flux-avg., $\sigma_\mu<0.5$     & $0.313 \pm 0.007$ & $-0.833 \pm 0.073$ & $-0.60 \pm 0.27$ \\
              &               & $(-0.3\%)$ & $(-2.5\%)$ & $(+12.2\%)$ \\
\midrule
\multirow{2}{*}{\makecell[l]{Union3 \\ {\scriptsize + Planck15 + DESI~DR2}}}
              & not flux-avg. & $0.327 \pm 0.009$ & $-0.693 \pm 0.091$ & $-0.97 \pm 0.32$ \\
              &               & $(-0.1\%)$ & $(-0.2\%)$ & $(+0.1\%)$ \\
\midrule \midrule
\multicolumn{5}{c}{$\mu(z)$~\citep{PaperI}} \\
\midrule
Dataset & flux-avg. mode & $\Omega_m$ & $w_0$ & $w_a$ \\
\midrule \midrule
\multirow{2}{*}{\makecell[l]{Pantheon \\ {\scriptsize + Planck15 + DESI~DR2}}}
              & not flux-avg. & $0.304 \pm 0.007$ & $-0.923 \pm 0.070$ & $-0.35 \pm 0.25$ \\
              & flux-avg.     & $0.302 \pm 0.007$ & $-0.944 \pm 0.072$ & $-0.29 \pm 0.25$ \\
\midrule
\multirow{2}{*}{\makecell[l]{Pantheon+ \\ {\scriptsize + Planck15 + DESI~DR2}}}
              & not flux-avg. & $0.311 \pm 0.006$ & $-0.863 \pm 0.056$ & $-0.48 \pm 0.22$ \\
              & flux-avg.     & $0.312 \pm 0.008$ & $-0.847 \pm 0.077$ & $-0.54 \pm 0.26$ \\
\midrule
\multirow{2}{*}{\makecell[l]{DES-Dovekie \\ {\scriptsize + Planck15 + DESI~DR2}}}
              & not flux-avg. & $0.313 \pm 0.006$ & $-0.834 \pm 0.057$ & $-0.59 \pm 0.24$ \\
              & flux-avg., $\sigma_\mu<0.5$ & $0.313 \pm 0.007$ & $-0.813 \pm 0.073$ & $-0.68 \pm 0.27$ \\
\midrule
\makecell[l]{Union3 \\ {\scriptsize + Planck15 + DESI~DR2}}
              & not flux-avg. & $0.327 \pm 0.009$ & $-0.692 \pm 0.091$ & $-0.97 \pm 0.32$ \\
\bottomrule
\end{tabular}
\end{table*}

\begin{figure*}[t]
    \centering
    \includegraphics[width=\linewidth]{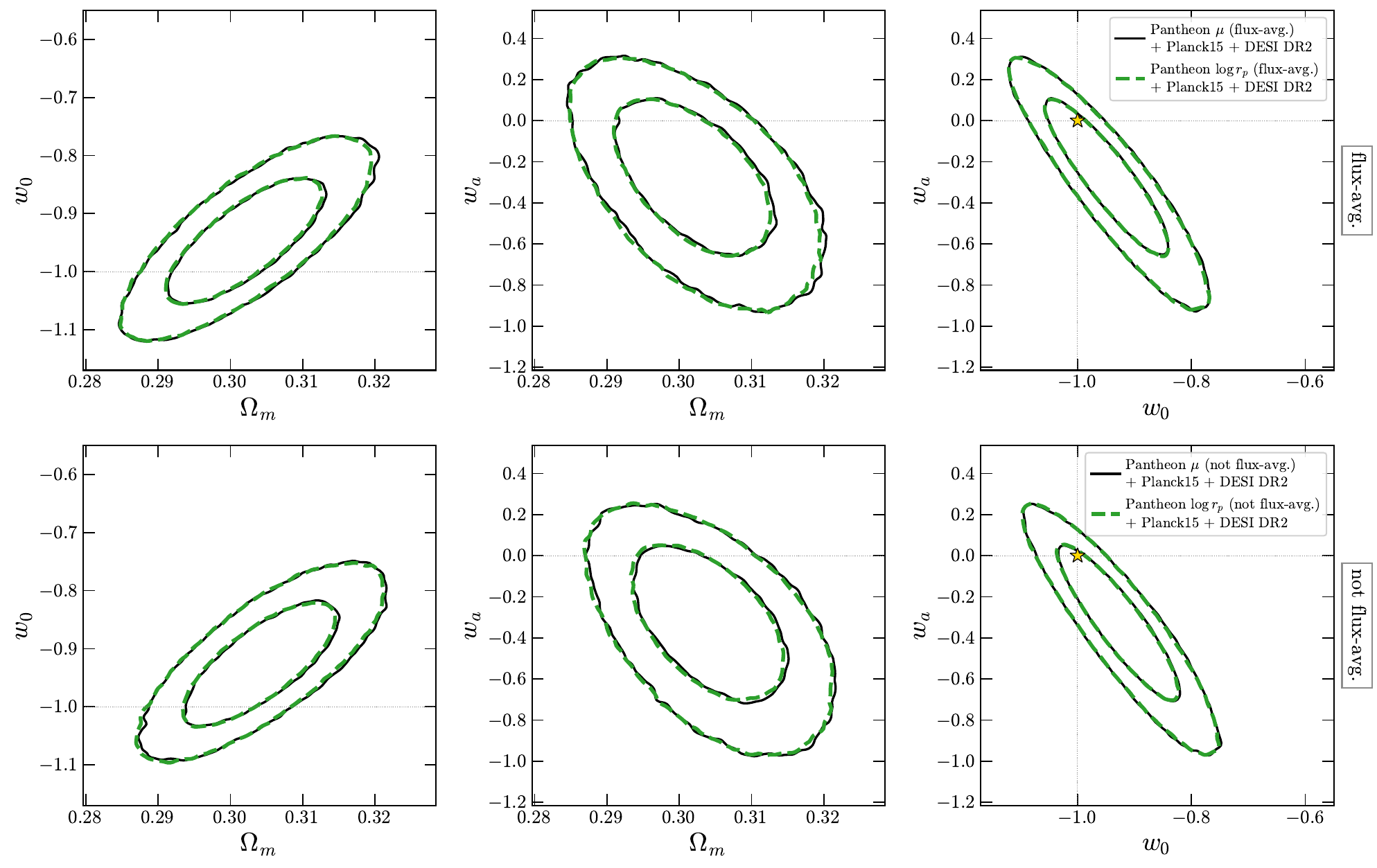}
    \caption{Joint posteriors on $(\Omega_m, w_0, w_a)$ in flat $w_0 w_a$CDM for Pantheon, comparing the closed-form $\log r_p$ analysis (this work, green dashed) with the direct distance-modulus analysis of \citep{PaperI} (black solid). {\it Top row:} Pantheon with flux averaging. {\it Bottom row:} Pantheon without flux averaging. Within each column, both panels share the same $y$-axis range so the small flux-averaging shift can be read off directly. All chains include the Planck~2015 CMB distance priors of \cite{WangDai2016} and DESI~DR2 BAO.}
    \label{fig:contour_Pantheon_CPL_6panel}
\end{figure*}

The $w_0 w_a$CDM comparison differs for not-flux-averaged and flux-averaged data. The four not-flux-averaged chains reproduce the $\mu$-based reference essentially exactly: fractional shifts of $\lesssim 0.2\%$ on both $\Omega_m$ and $w_0$ and $\lesssim 1.5\%$ on $w_a$ for every dataset, with 68\% C.L.\ ranges agreeing at the percent level. The flux-averaged chains, in contrast, show dataset-dependent residuals: shifts of $\lesssim 0.2\%$ for Pantheon, but $3.1\%$ in $w_0$ / $18.9\%$ in $w_a$ for Pantheon+ and $2.5\%$ / $12.2\%$ for DES-Dovekie, with 68\% C.L.\ ranges inflated by $\lesssim 5\%$. This is because the two pipelines do not implement flux averaging identically: the $\log r_p$-side estimator and the $\mu$-side MCMC apply the luminosity-weighting step at different cosmologies (Sec.~\ref{subsec:fa_newton}), so the binned data feeding the downstream likelihoods differ slightly. As an example, the full joint $(\Omega_m, w_0, w_a)$ posterior comparison is shown in Fig.~\ref{fig:contour_Pantheon_CPL_6panel} for Pantheon; all three two-dimensional contours and all three one-dimensional marginals overlap between the compressed and full analyses, in both flux-averaged and not-flux-averaged modes.

\subsubsection{Model-independent dark energy measurement $X(z)$CDM}
For the model-independent dark-energy reconstruction, we sample $(\omega_m, \omega_b, h, X_1, \ldots, X_5)$ at the five $z$ knots $\{1/3, 2/3, 1, 4/3, 2.33\}$, again combining the $\log r_p$ likelihood with the same CMB+BAO priors. Results are listed in Table~\ref{tab:Xz_from_rp}. Note that the $\omega_b$ is constrained by CMB only so we do not list it for comparison.

\begin{table*}[t]
\centering
\caption{Non-parametric $X(z) = \rho_{\rm DE}(z)/\rho_{\rm DE}(0)$ knot values from the closed-form $\log r_p(z)$ posterior plus the Planck~2015 CMB distance priors of \cite{WangDai2016} and DESI~DR2 BAO (top section), compared with the corresponding full distance--modulus chains (bottom section). Not-flux-averaged $\mu(z)$ entries are the original Paper~I~\cite{PaperI} chains; flux-averaged $\mu(z)$ entries for Pantheon+ and DES-Dovekie are recomputed using the hybrid binning of Sec.~\ref{subsec:fa_newton}. The five $X_k$ columns give the dark-energy density at the knot redshifts $z_k \in \{1/3,\, 2/3,\, 1,\, 4/3,\, 2.33\}$. In the top section each $\log r_p$ row is followed by a subrow giving the fractional difference $\Delta\theta/\theta\equiv \left(\theta^{\log r_p} - \theta^{\mu(z)}\right)/\theta^{\mu(z)}$. We use four chains, each uses $2\times 10^5 \times 4$ samples ($\sim$6\,min per dataset).}
\label{tab:Xz_from_rp}
\renewcommand{\arraystretch}{1.1}
\scriptsize
\setlength{\tabcolsep}{2pt}
\resizebox{\textwidth}{!}{%
\begin{tabular}{@{}llccccccc@{}}
\toprule
\multicolumn{9}{c}{$\log r_p$ (this work)} \\
\midrule
Dataset & flux-avg. mode & $\Omega_m$ & $H_0$ & $X(\tfrac{1}{3})$ & $X(\tfrac{2}{3})$ & $X(1)$ & $X(\tfrac{4}{3})$ & $X(2.33)$ \\
\midrule \midrule
\multirow{4}{*}{Pantheon}
              & not flux-avg. & $0.304 \pm 0.007$ & $68.54 \pm 0.74$ & $1.00 \pm 0.04$ & $1.12 \pm 0.07$ & $0.91 \pm 0.10$ & $0.84 \pm 0.16$ & $0.72 \pm 0.31$ \\
              &               & $(-0.02\%)$ & $(<0.01\%)$ & $(-0.04\%)$ & $(+0.03\%)$ & $(<0.01\%)$ & $(+0.1\%)$ & $(+0.3\%)$ \\
              & flux-avg.     & $0.302 \pm 0.007$ & $68.78 \pm 0.76$ & $0.99 \pm 0.04$ & $1.11 \pm 0.07$ & $0.90 \pm 0.10$ & $0.83 \pm 0.16$ & $0.71 \pm 0.30$ \\
              &               & $(-0.01\%)$ & $(<0.01\%)$ & $(+0.03\%)$ & $(-0.1\%)$ & $(-0.1\%)$ & $(+0.01\%)$ & $(+0.2\%)$ \\
\midrule
\multirow{4}{*}{Pantheon+}
              & not flux-avg. & $0.312 \pm 0.006$ & $67.64 \pm 0.60$ & $1.05 \pm 0.03$ & $1.19 \pm 0.07$ & $0.91 \pm 0.10$ & $0.86 \pm 0.17$ & $0.75 \pm 0.32$ \\
              &               & $(-0.03\%)$ & $(+0.02\%)$ & $(-0.1\%)$ & $(-0.1\%)$ & $(+0.03\%)$ & $(-0.1\%)$ & $(-0.3\%)$ \\
              & flux-avg.     & $0.311 \pm 0.008$ & $67.76 \pm 0.82$ & $1.03 \pm 0.05$ & $1.22 \pm 0.08$ & $0.91 \pm 0.11$ & $0.87 \pm 0.17$ & $0.73 \pm 0.32$ \\
              &               & $(-0.4\%)$ & $(+0.2\%)$ & $(-0.8\%)$ & $(-0.5\%)$ & $(-0.7\%)$ & $(-0.4\%)$ & $(-0.9\%)$ \\
\midrule
\multirow{4}{*}{DES-Dovekie}
              & not flux-avg. & $0.312 \pm 0.006$ & $67.69 \pm 0.57$ & $1.07 \pm 0.03$ & $1.09 \pm 0.06$ & $1.01 \pm 0.11$ & $0.85 \pm 0.16$ & $0.76 \pm 0.32$ \\
              &               & $(+0.02\%)$ & $(-0.01\%)$ & $(+0.04\%)$ & $(+0.1\%)$ & $(-0.01\%)$ & $(+0.1\%)$ & $(+0.2\%)$ \\
              & flux-avg.($\sigma_\mu < 0.5$)    & $0.310 \pm 0.007$ & $67.83 \pm 0.75$ & $1.05 \pm 0.05$ & $1.12 \pm 0.07$ & $0.97 \pm 0.11$ & $0.85 \pm 0.17$ & $0.75 \pm 0.32$ \\
              &               & $(-0.2\%)$ & $(+0.1\%)$ & $(-0.3\%)$ & $(-0.4\%)$ & $(-0.1\%)$ & $(-0.2\%)$ & $(+0.1\%)$ \\
\midrule
\multirow{2}{*}{Union3} & not flux-avg. & $0.324 \pm 0.009$ & $66.42 \pm 0.88$ & $1.15 \pm 0.06$ & $1.21 \pm 0.08$ & $1.01 \pm 0.12$ & $0.92 \pm 0.18$ & $0.80 \pm 0.34$ \\
              &               & $(-0.1\%)$ & $(+0.03\%)$ & $(-0.1\%)$ & $(-0.1\%)$ & $(-0.1\%)$ & $(-0.1\%)$ & $(+0.04\%)$ \\
\midrule \midrule
\multicolumn{9}{c}{$\mu(z)$~\citep{PaperI}} \\
\midrule
Dataset & flux-avg. mode & $\Omega_m$ & $H_0$ & $X(\tfrac{1}{3})$ & $X(\tfrac{2}{3})$ & $X(1)$ & $X(\tfrac{4}{3})$ & $X(2.33)$ \\
\midrule \midrule
\multirow{2}{*}{Pantheon}
              & not flux-avg. & $0.304 \pm 0.007$ & $68.53 \pm 0.74$ & $1.00 \pm 0.04$ & $1.12 \pm 0.07$ & $0.91 \pm 0.10$ & $0.84 \pm 0.16$ & $0.72 \pm 0.31$ \\
              & flux-avg.     & $0.302 \pm 0.007$ & $68.79 \pm 0.77$ & $0.99 \pm 0.04$ & $1.11 \pm 0.07$ & $0.90 \pm 0.10$ & $0.83 \pm 0.16$ & $0.71 \pm 0.30$ \\
\midrule
\multirow{2}{*}{Pantheon+}
              & not flux-avg. & $0.312 \pm 0.006$ & $67.63 \pm 0.60$ & $1.05 \pm 0.03$ & $1.19 \pm 0.07$ & $0.91 \pm 0.10$ & $0.86 \pm 0.17$ & $0.75 \pm 0.32$ \\
              & flux-avg.     & $0.312 \pm 0.008$ & $67.63 \pm 0.82$ & $1.04 \pm 0.05$ & $1.23 \pm 0.09$ & $0.92 \pm 0.11$ & $0.87 \pm 0.17$ & $0.74 \pm 0.32$ \\
\midrule
\multirow{2}{*}{DES-Dovekie}
              & not flux-avg. & $0.312 \pm 0.006$ & $67.69 \pm 0.57$ & $1.07 \pm 0.03$ & $1.09 \pm 0.06$ & $1.01 \pm 0.11$ & $0.85 \pm 0.16$ & $0.76 \pm 0.32$ \\
              & flux-avg.($\sigma_\mu < 0.5$)  & $0.311 \pm 0.007$ & $67.78 \pm 0.75$ & $1.05 \pm 0.05$ & $1.13 \pm 0.07$ & $0.97 \pm 0.11$ & $0.85 \pm 0.17$ & $0.75 \pm 0.32$ \\
\midrule
Union3        & not flux-avg. & $0.324 \pm 0.009$ & $66.41 \pm 0.88$ & $1.15 \pm 0.06$ & $1.21 \pm 0.08$ & $1.01 \pm 0.12$ & $0.92 \pm 0.18$ & $0.80 \pm 0.33$ \\
\bottomrule
\end{tabular}%
}
\end{table*}

\begin{figure*}[t]
    \centering
    \includegraphics[width=\linewidth]{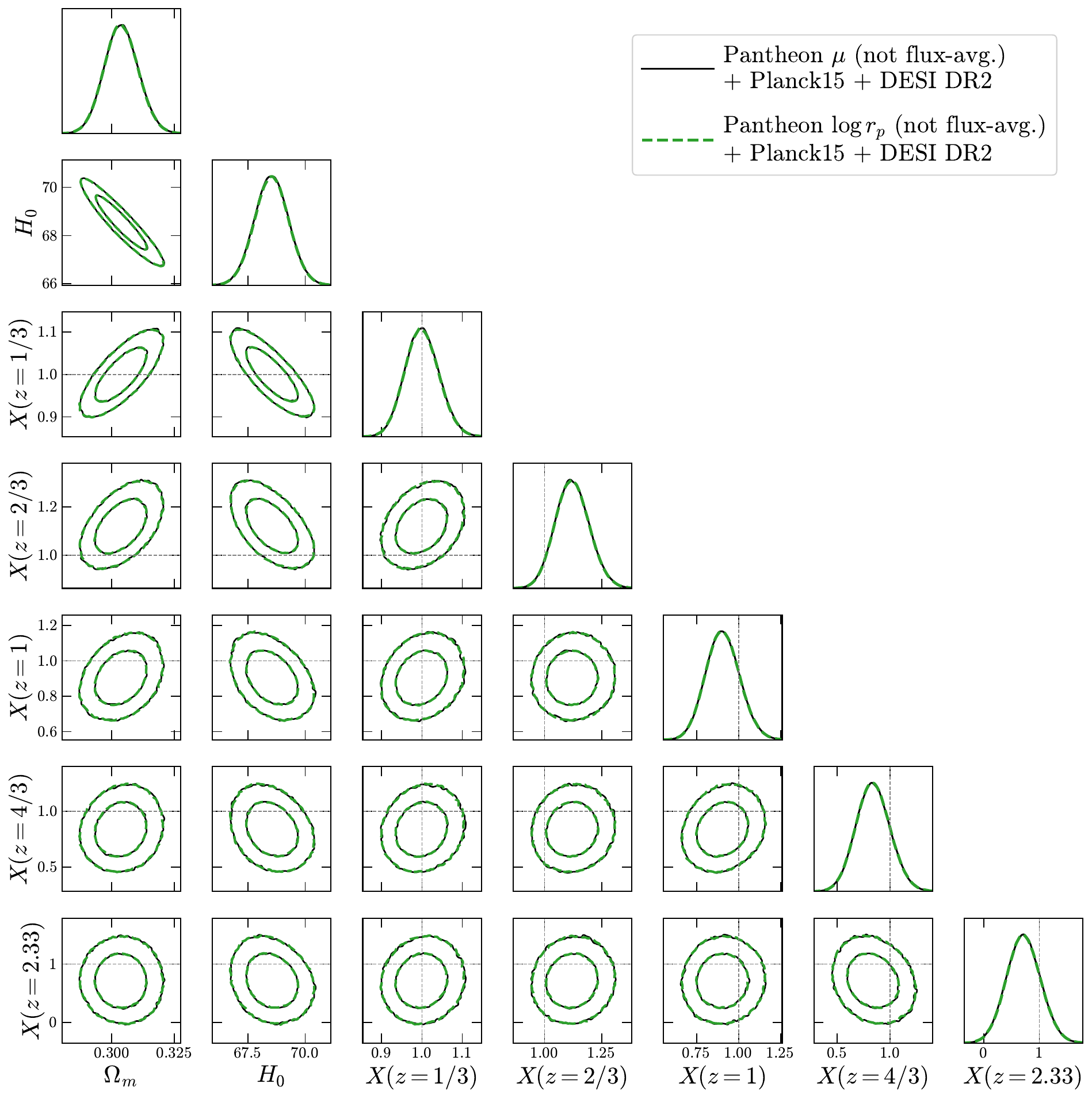}
    \caption{Joint posteriors on $(\Omega_m, H_0, X_1, \ldots, X_5)$ from the non-parametric $X(z)$ analysis of the Pantheon dataset (not flux-averaged), comparing the closed-form $\log r_p$ pipeline (this work, green dashed) with the direct distance-modulus pipeline of \citep{PaperI} (black solid). Diagonal panels show the 1D marginals; lower-triangle panels show the 68\% and 95\% confidence contours. Grey dashed reference lines mark the $\Lambda$CDM value $X = 1$. Both chains include the Planck~2015 CMB distance priors in \cite{WangDai2016} and DESI~DR2 BAO.}
    \label{fig:corner_Pantheon_Xz_noFA}
\end{figure*}

\begin{figure*}[t]
    \centering
    \includegraphics[width=\linewidth]{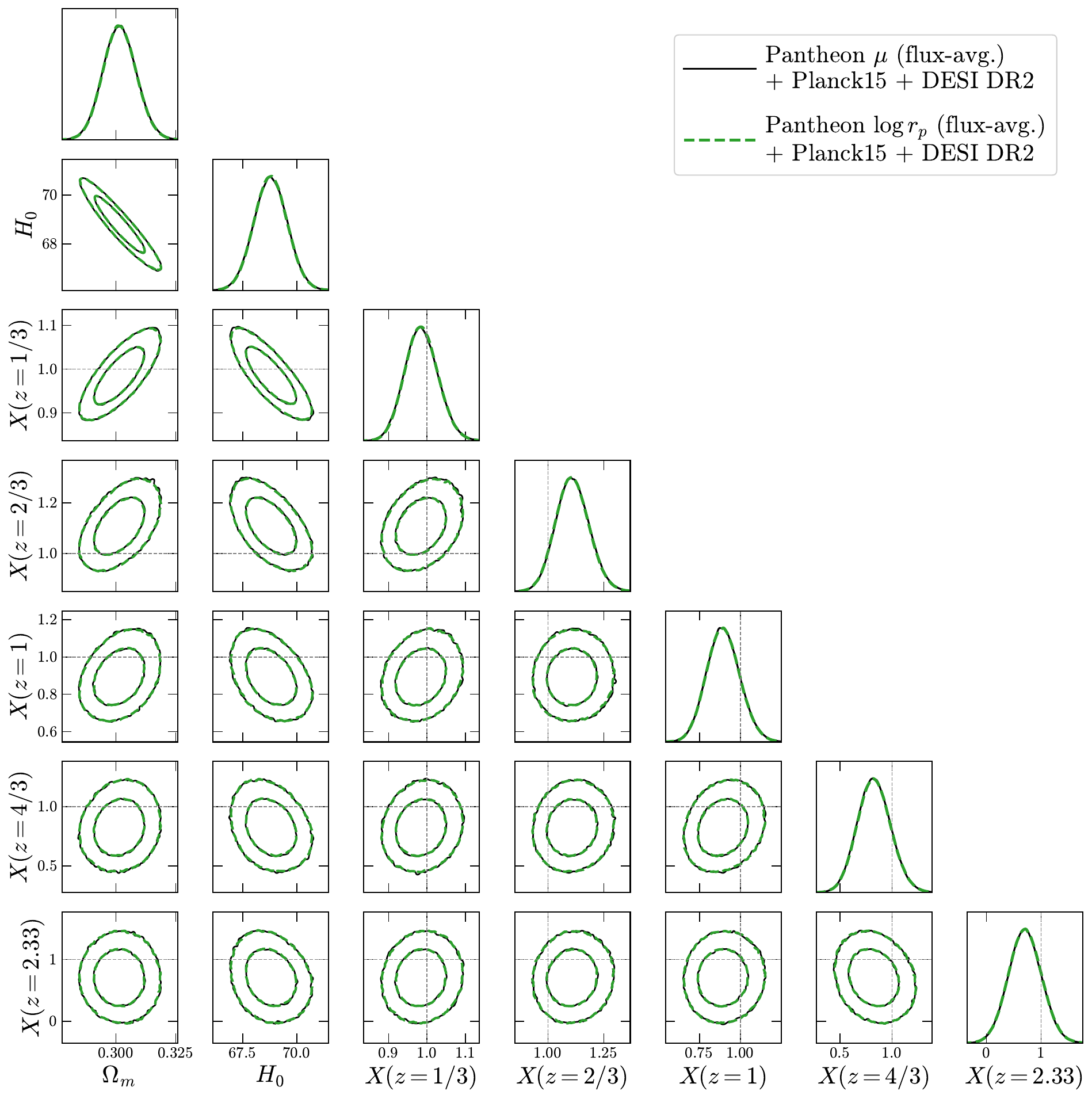}
    \caption{Same as Fig.~\ref{fig:corner_Pantheon_Xz_noFA} but for the flux-averaged Pantheon dataset.}
    \label{fig:corner_Pantheon_Xz_FA}
\end{figure*}

The non-parametric $X(z)$ reconstructions show the same dichotomy between flux-averaged and not-flux-averaged chains. Not-flux-averaged chains match the $\mu$-based references essentially exactly: $|\Delta X_k|/X_k \lesssim 0.3\%$ at every knot for all four datasets. Flux-averaged chains again show a small residual: Pantheon+ with flux averaging reaches up to $0.9\%$ at $X(2.33)$ and DES-Dovekie with flux averaging up to $0.4\%$ at $X(2/3)$, mirroring their $w_0 w_a$CDM behavior, while Pantheon with flux averaging remains at $\lesssim 0.2\%$. The full $(\Omega_m, H_0, X_1, \ldots, X_5)$ joint posteriors for Pantheon are shown in Fig.~\ref{fig:corner_Pantheon_Xz_noFA} (not-flux-averaged) and Fig.~\ref{fig:corner_Pantheon_Xz_FA} (flux-averaged). All 1D marginals and lower-triangle 2D contours overlap between the compressed and full analyses.

As an additional cross-check we evaluate the Figure-of-Merit ($\mathrm{FoM} \equiv 1/\sqrt{\det \mathbf{C}_{\bm\theta}}$, Sec.~\ref{sec:intro}) on chains, both on the dark-energy subspace and on the full sampled parameter space. For $w_0 w_a$CDM the dark-energy $\mathrm{FoM}(w_0, w_a)$ is preserved between the two pipelines to within $\sim 2\%$ across all four datasets in both flux-averaging modes (ratios $0.998$--$1.021$), while the full five-parameter $\mathrm{FoM}(\Omega_m, \omega_b, H_0, w_0, w_a)$ is preserved to within $\sim 6\%$, with the $\log r_p$-side posterior systematically slightly tighter (ratios $1.045$--$1.063$). The non-parametric $X(z)$ chains follow the same qualitative pattern. The variation across datasets is at the percent level.

To summarize the computational cost across the three cosmologies and seven dataset/flux-averaged configurations, the closed-form analytical $\log r_p$ extraction (App.~\ref{app:linear_gaussian}) takes \textless\,$0.1$\,s per dataset, and the downstream cosmology MCMC takes $1$--$12$\,min per chain: roughly $1$\,min for flat $\Lambda$CDM ($\Omega_m$ alone), $8$--$12$\,min for $w_0 w_a$CDM, and $5$--$7$\,min for the non-parametric $X(z)$. The complete reproduction of Paper~I's three-cosmology constraint set (21 chains in total, with sample sizes equal to or larger than Paper~I) finishes in $\sim 1$\,h of wall clock time when the three pipelines are run in parallel, compared to $\sim 20$\,h for the original full-$\mu$ MCMC; each Paper~I full-$\mu$ CPL chain takes $\sim 4$\,h for the flux-averaged Pantheon+ ($230$\,min) and DES-Dovekie ($247$\,min) datasets, against $\sim 10$\,min per chain on the $\log r_p$ side, a factor of $\sim 25$ speedup per chain. The $\log r_p$ posterior compresses ${\sim}1600$ distance moduli into an $11 \times 11$ Gaussian, eliminating the $N_{\rm SN} \times N_{\rm SN}$ matrix operations at each downstream MCMC step.

\section{Discussion and Conclusions}
\label{sec:discussion}
We have presented a model-independent and lossless compression of the SNe~Ia distance--redshift information into a Gaussian posterior on $\log r_p(z)$ knots. The pipeline has two parts: a simple estimator completes the efficient compression for each dataset (closed-form BLUE for the not-flux-averaged data, two-step Newton iteration for the flux-averaged data), and a downstream cosmological MCMC that uses $(\hat{\mathbf y}, \boldsymbol{\Sigma}_y)$ as a plug-in $11$-dimensional Gaussian likelihood and runs in minutes per chain regardless of which cosmological model is being tested.

Across all $21$ chains we ran (seven dataset/flux-averaged configurations $\times$ three cosmological models: $\Lambda$CDM, $w_0 w_a$CDM, and a non-parametric $X(z)$), the downstream parameter contours and figures of merit reproduce the corresponding direct distance-modulus analyses of \citep{PaperI} to within statistical sampling noise for the not-flux-averaged data. The flux-averaged chains show small dataset-dependent residuals (Sec.~\ref{subsec:cosmo_from_rp}) that trace back to a difference in how the two pipelines implement the flux-averaging step rather than to information loss in the compression. The compression is therefore lossless in the operational sense defined in Sec.~\ref{sec:intro}: a downstream cosmological analysis cannot tell whether the SNe~Ia information was supplied as the full $(\bm\mu, \mathbf C_\mu)$ or as the eleven $\log r_p$ values with their covariance matrix.

The practical implication grows with the data volume of the next decade. Roman, Euclid, and LSST will deliver SNe~Ia samples one to three orders of magnitude larger than current compilations. For the direct distance-modulus analysis the per-likelihood-evaluation cost scales with $N_{\rm SN}$ through the $N_{\rm SN} \times N_{\rm SN}$ covariance, so the chain wall clock time grows accordingly, and exploring different dark-energy parameterizations becomes much more expensive. The compressed pipeline behaves very differently: the analytic estimator runs as a one-shot precompute and produces the same $11 \times 11$ Gaussian regardless of $N_{\rm SN}$; the downstream MCMC then operates on a fixed-dimensional likelihood whose per-step cost is set by the number of knots, not by $N_{\rm SN}$. Increasing the number of knots to track the additional SN density that Roman and Euclid will deliver at $z \gtrsim 0.8$ is cheap: the analytic-estimator cost is $\mathcal{O}(K\times N_{\rm SN})$ and the downstream per-step likelihood cost is $\mathcal{O}(K^2)$, so going from $K=11$ to, say, $K=15$ knots adds only a constant-factor overhead to an already sub-second extraction and minute-scale MCMC.

Two concrete applications follow from this scaling. First, the same compressed product $(\hat{\mathbf y}, \boldsymbol{\Sigma}_y)$ serves as a portable, deterministic, bit-reproducible SN Ia distance prior for joint analyses with CMB, BAO, and other probes, putting SNe~Ia on the same operational footing already enjoyed by the CMB through the shift parameters $(R, l_a, \omega_b)$ of \cite{WangMukherjee2007, WangDai2016}. Second, any dark-energy model that has not yet been written down can be tested by feeding $(\hat{\mathbf y}, \boldsymbol{\Sigma}_y)$ into a downstream MCMC with the new theory while the upstream compression cost is paid once.

\paragraph{Code and data availability.} The compressed $(\hat{\bm y}, \boldsymbol{\Sigma}_y)$ data products are tabulated in Appendix~\ref{app:compressed_data}. The pipeline code will be made publicly available upon the publication of this paper.

\acknowledgments

ZW thanks Henry S Grasshorn Gebhardt, Utkarsh Giri, and Fei Ge for useful discussion. We gratefully acknowledge support from NASA Grant \#80NSSC24M0021, ``Project Infrastructure for the Roman Galaxy Redshift Survey'', and NASA ROSES Grant 12-EUCLID11-0004.

\appendix

\section{Mathematical Background}
\label{app:logrp_closed_form}

This appendix reviews two ingredients underlying the main-text estimators: the optimal Gaussian-posterior estimator for linear-Gaussian models (Sec.~\ref{app:linear_gaussian}) and Newton's method for non-quadratic objectives (Sec.~\ref{app:newton_review}).

\subsection{Linear-Gaussian model and the BLUE estimator}
\label{app:linear_gaussian}

Consider a linear forward model
\begin{equation}
    \mathbf{d} = \mathbf{A}\,\boldsymbol{\theta} + \boldsymbol{\epsilon},
\end{equation}
where $\mathbf{d} \in \mathbb{R}^N$ is the observed \emph{data}, $\boldsymbol{\theta} \in \mathbb{R}^K$ is the vector of \emph{parameters} to be inferred, $\mathbf{A} \in \mathbb{R}^{N\times K}$ is the (parameter-independent) \emph{linear forward model} that maps parameters to expected data, and $\boldsymbol{\epsilon} \sim \mathcal{N}(0, \mathbf{C})$ is the \emph{Gaussian noise} with known covariance $\mathbf{C}$. The negative log-likelihood,
\begin{equation}
    -2\ln\mathcal{L}(\boldsymbol{\theta}) = (\mathbf{d} - \mathbf{A}\boldsymbol{\theta})^T\,\mathbf{C}^{-1}\,(\mathbf{d} - \mathbf{A}\boldsymbol{\theta}),
\end{equation}
is quadratic in $\boldsymbol{\theta}$. Setting its gradient to zero,
\begin{equation}
    -\tfrac{1}{2}\,\frac{\partial(-2\ln\mathcal{L})}{\partial \boldsymbol{\theta}} = \mathbf{A}^T \mathbf{C}^{-1}\,(\mathbf{d} - \mathbf{A}\boldsymbol{\theta}) = 0,
\end{equation}
gives the best linear unbiased estimator (BLUE),
\begin{equation}
    \hat{\boldsymbol{\theta}} = (\mathbf{A}^T \mathbf{C}^{-1} \mathbf{A})^{-1}\,\mathbf{A}^T \mathbf{C}^{-1}\,\mathbf{d}.
    \label{eq:appA_blue}
\end{equation}
Expanding $-2\ln\mathcal{L}$ around $\hat{\boldsymbol{\theta}}$ identifies the Fisher information $\mathbf{F} \equiv \mathbf{A}^T \mathbf{C}^{-1} \mathbf{A}$ as the Hessian of $-\ln\mathcal{L}$, so under a flat prior the posterior is exactly Gaussian, $p(\boldsymbol{\theta}\,|\,\mathbf{d}) = \mathcal{N}(\hat{\boldsymbol{\theta}}, \boldsymbol{\Sigma})$ with $\boldsymbol{\Sigma} = \mathbf{F}^{-1}$.

For the SNe~Ia application the model carries an additional unknown global offset $\mathcal{M}$, i.e.\ $\mathbf{d} = \mathbf{A}\boldsymbol{\theta} + \mathcal{M}\,\mathbf{1} + \boldsymbol{\epsilon}$, that we marginalise out analytically under a flat prior. Setting $\partial(-2\ln\mathcal{L})/\partial \mathcal{M} = 0$ at fixed $\boldsymbol{\theta}$ gives
\begin{equation}
    \hat{\mathcal{M}}(\boldsymbol{\theta}) = \frac{\mathbf{1}^T\mathbf{C}^{-1}(\mathbf{d}-\mathbf{A}\boldsymbol{\theta})}{\mathbf{1}^T\mathbf{C}^{-1}\mathbf{1}},
\end{equation}
and substituting back yields a marginal likelihood that is again quadratic in $\boldsymbol{\theta}$ but with $\mathbf{C}^{-1}$ replaced by the rank-$(N-1)$ projection
\begin{equation}
    \mathbf{M} = \mathbf{C}^{-1} - \frac{\mathbf{C}^{-1}\mathbf{1}\mathbf{1}^T \mathbf{C}^{-1}}{\mathbf{1}^T \mathbf{C}^{-1}\mathbf{1}},
    \label{eq:appA_M_projection}
\end{equation}
which annihilates the data subspace parallel to $\mathbf{1}$. Re-running the gradient-zero argument above with $\mathbf{C}^{-1} \to \mathbf{M}$ gives the $\mathcal{M}$-marginalised BLUE used in Eqs.~(\ref{eq:closed_form_main}) and (\ref{eq:M_projection}) of Sec.~\ref{subsec:closed_form}, with $\boldsymbol{\theta} = \bm{y}$, $\mathbf{d} = \boldsymbol{\mu}_{\rm obs} - \boldsymbol{\nu}$.

\subsection{Newton's method and the Hessian}
\label{app:newton_review}

For a smooth objective $f(\boldsymbol{\theta})$, Newton's method approximates $f$ near the current iterate $\boldsymbol{\theta}^{(j)}$ by its second-order Taylor expansion,
\begin{equation}
    f(\boldsymbol{\theta}^{(j)} + \Delta\boldsymbol{\theta}) \approx f(\boldsymbol{\theta}^{(j)}) + (\nabla f)^T \Delta\boldsymbol{\theta} + \tfrac{1}{2}\,\Delta\boldsymbol{\theta}^T \mathbf{H}\,\Delta\boldsymbol{\theta},
\end{equation}
where the Hessian $\mathbf{H} \equiv \nabla\nabla^T f$ is the matrix of second partial derivatives. Setting the derivative with respect to $\Delta\boldsymbol{\theta}$ to zero gives $\Delta\boldsymbol{\theta} = -\mathbf{H}^{-1}\,\nabla f$, so each Newton step is
\begin{equation}
    \boldsymbol{\theta}^{(j+1)} = \boldsymbol{\theta}^{(j)} - \mathbf{H}^{-1}(\boldsymbol{\theta}^{(j)})\,\nabla f(\boldsymbol{\theta}^{(j)}).
    \label{eq:appA_newton_step}
\end{equation}
The Hessian rescales the gradient by the local curvature, so flat directions take large steps and steep directions take small ones. Near a local minimum the iteration converges quadratically, much faster than the linear convergence of plain gradient descent. For $f = -2\ln\mathcal{L}$ with an approximately Gaussian posterior at the optimum $\boldsymbol{\theta}^*$, the Hessian is twice the inverse posterior covariance,
\begin{equation}
    \boldsymbol{\Sigma}^* = \tfrac{1}{2}\mathbf{H}^{-1}(\boldsymbol{\theta}^*),
    \label{eq:appA_hessian_cov}
\end{equation}
which is the relation used in Sec.~\ref{subsec:fa_newton} to obtain $\bar{\boldsymbol{\Sigma}}$ from the Hessian of Eq.~(\ref{eq:logL_FA}). In our pipeline, $\nabla f$ and $\mathbf{H}$ are evaluated at each step by automatic differentiation through the full forward model.


\section{Compressed \texorpdfstring{$\log r_p$}{log rp} Data Products}
\label{app:compressed_data}

Tables~\ref{tab:logrp_Pantheon}--\ref{tab:logrp_Union3} report the compressed data product $(\hat{\mathbf y}, \boldsymbol{\Sigma}_y)$ for the four SNe~Ia compilations (Pantheon, Pantheon+, DES-Dovekie, Union3). Each table consists of two halves: the not-flux-averaged BLUE estimator of Sec.~\ref{subsec:closed_form}, and the flux-averaged iterative-Newton estimator of Sec.~\ref{subsec:fa_newton}; Union3 is pre-binned at the catalogue level and only compressed without flux averaging. All values are reported with the FLRW kinematic anchor $\log r_p(z_0 = 0) = 0$ fixed.
Mean values are quoted to five significant digits and covariance entries to four significant digits. The knot redshifts $z_k$ are listed in Table~\ref{tab:knots} and reproduced in each table for convenience. These tables are intended as the self-contained, archive-quality SNe~Ia distance prior of this work. Downstream cosmological pipelines can plug $(\hat{\bm y}, \boldsymbol{\Sigma}_y)$ directly into a Gaussian likelihood and recover the constraints reported in Sec.~\ref{sec:results} to within Monte-Carlo noise on the chain.


\begin{table}[p]
\centering
\caption{Compressed $\log r_p$ data product for Pantheon.}
\label{tab:logrp_Pantheon}
\setlength{\tabcolsep}{3pt}

\centerline{\textbf{Pantheon (not flux-avg.)}}
\smallskip

\resizebox{\textwidth}{!}{
\begin{tabular}{l|ccccccccccc}
\toprule
 & $k{=}1$ & $k{=}2$ & $k{=}3$ & $k{=}4$ & $k{=}5$ & $k{=}6$ & $k{=}7$ & $k{=}8$ & $k{=}9$ & $k{=}10$ & $k{=}11$ \\
 & $z{=}0.05$ & $z{=}0.15$ & $z{=}0.30$ & $z{=}0.40$ & $z{=}0.50$ & $z{=}0.60$ & $z{=}0.70$ & $z{=}0.80$ & $z{=}1.00$ & $z{=}1.60$ & $z{=}2.27$ \\
\midrule
$10^{2} \log r_p$ & $1.6699$ & $2.4215$ & $1.3705$ & $3.0354$ & $2.1871$ & $2.6020$ & $1.6244$ & $2.1486$ & $3.0401$ & $-1.4147$ & $-5.7041$ \\
\midrule
$10^{4} \Sigma_{kl},\,k{=}1$ & $2.466$ &  &  &  &  &  &  &  &  &  &  \\
$k{=}2$ & $2.380$ & $3.198$ &  &  &  &  &  &  &  &  &  \\
$k{=}3$ & $2.445$ & $2.810$ & $3.051$ &  &  &  &  &  &  &  &  \\
$k{=}4$ & $2.165$ & $2.876$ & $2.699$ & $3.255$ &  &  &  &  &  &  &  \\
$k{=}5$ & $2.330$ & $2.687$ & $2.705$ & $2.696$ & $3.749$ &  &  &  &  &  &  \\
$k{=}6$ & $2.449$ & $3.059$ & $2.905$ & $3.003$ & $2.978$ & $4.551$ &  &  &  &  &  \\
$k{=}7$ & $2.400$ & $2.926$ & $2.833$ & $2.887$ & $3.085$ & $3.437$ & $4.955$ &  &  &  &  \\
$k{=}8$ & $2.293$ & $2.856$ & $2.760$ & $2.839$ & $2.960$ & $3.222$ & $3.490$ & $4.442$ &  &  &  \\
$k{=}9$ & $2.012$ & $2.228$ & $2.252$ & $2.233$ & $2.637$ & $2.781$ & $3.078$ & $2.805$ & $5.511$ &  &  \\
$k{=}10$ & $1.760$ & $1.728$ & $1.567$ & $1.494$ & $1.905$ & $2.411$ & $2.005$ & $3.387$ & $2.359$ & $22.79$ &  \\
$k{=}11$ & $1.484$ & $1.127$ & $1.244$ & $0.9116$ & $1.709$ & $1.580$ & $2.755$ & $0.6574$ & $6.165$ & $3.064$ & $125.1$ \\
\bottomrule
\end{tabular}}

\vspace{2ex}

\centerline{\textbf{Pantheon (flux-avg.)}}
\smallskip

\resizebox{\textwidth}{!}{
\begin{tabular}{l|ccccccccccc}
\toprule
 & $k{=}1$ & $k{=}2$ & $k{=}3$ & $k{=}4$ & $k{=}5$ & $k{=}6$ & $k{=}7$ & $k{=}8$ & $k{=}9$ & $k{=}10$ & $k{=}11$ \\
 & $z{=}0.05$ & $z{=}0.15$ & $z{=}0.30$ & $z{=}0.40$ & $z{=}0.50$ & $z{=}0.60$ & $z{=}0.70$ & $z{=}0.80$ & $z{=}1.00$ & $z{=}1.60$ & $z{=}2.27$ \\
\midrule
$10^{2} \log r_p$ & $2.0306$ & $3.1632$ & $2.0042$ & $3.4568$ & $2.9392$ & $3.6122$ & $2.7396$ & $3.1216$ & $3.7599$ & $0.25188$ & $-5.7477$ \\
\midrule
$10^{4} \Sigma_{kl},\,k{=}1$ & $2.599$ &  &  &  &  &  &  &  &  &  &  \\
$k{=}2$ & $2.488$ & $3.346$ &  &  &  &  &  &  &  &  &  \\
$k{=}3$ & $2.565$ & $2.939$ & $3.208$ &  &  &  &  &  &  &  &  \\
$k{=}4$ & $2.256$ & $3.007$ & $2.825$ & $3.446$ &  &  &  &  &  &  &  \\
$k{=}5$ & $2.433$ & $2.810$ & $2.826$ & $2.814$ & $3.962$ &  &  &  &  &  &  \\
$k{=}6$ & $2.543$ & $3.182$ & $3.024$ & $3.127$ & $3.063$ & $4.794$ &  &  &  &  &  \\
$k{=}7$ & $2.507$ & $3.055$ & $2.961$ & $3.005$ & $3.219$ & $3.549$ & $5.244$ &  &  &  &  \\
$k{=}8$ & $2.394$ & $2.979$ & $2.880$ & $2.957$ & $3.083$ & $3.341$ & $3.636$ & $4.636$ &  &  &  \\
$k{=}9$ & $2.095$ & $2.320$ & $2.349$ & $2.321$ & $2.740$ & $2.861$ & $3.216$ & $2.856$ & $5.979$ &  &  \\
$k{=}10$ & $1.826$ & $1.797$ & $1.639$ & $1.554$ & $1.966$ & $2.473$ & $2.009$ & $3.567$ & $2.136$ & $25.67$ &  \\
$k{=}11$ & $1.553$ & $1.192$ & $1.314$ & $0.9591$ & $1.781$ & $1.606$ & $2.862$ & $0.7416$ & $6.469$ & $2.788$ & $129.4$ \\
\bottomrule
\end{tabular}}

\end{table}

\begin{table}[p]
\centering
\caption{Compressed $\log r_p$ data product for Pantheon+.}
\label{tab:logrp_PantheonPlus}
\setlength{\tabcolsep}{3pt}

\centerline{\textbf{Pantheon+ (not flux-avg.)}}
\smallskip

\resizebox{\textwidth}{!}{
\begin{tabular}{l|ccccccccccc}
\toprule
 & $k{=}1$ & $k{=}2$ & $k{=}3$ & $k{=}4$ & $k{=}5$ & $k{=}6$ & $k{=}7$ & $k{=}8$ & $k{=}9$ & $k{=}10$ & $k{=}11$ \\
 & $z{=}0.05$ & $z{=}0.15$ & $z{=}0.30$ & $z{=}0.40$ & $z{=}0.50$ & $z{=}0.60$ & $z{=}0.70$ & $z{=}0.80$ & $z{=}1.00$ & $z{=}1.60$ & $z{=}2.27$ \\
\midrule
$10^{2} \log r_p$ & $-0.77829$ & $-1.5318$ & $-0.89756$ & $-1.4297$ & $-2.6729$ & $-2.6291$ & $-2.5855$ & $-4.4896$ & $5.1462$ & $-4.4898$ & $0.81718$ \\
\midrule
$10^{4} \Sigma_{kl},\,k{=}1$ & $1.358$ &  &  &  &  &  &  &  &  &  &  \\
$k{=}2$ & $1.137$ & $1.277$ &  &  &  &  &  &  &  &  &  \\
$k{=}3$ & $1.248$ & $1.150$ & $1.478$ &  &  &  &  &  &  &  &  \\
$k{=}4$ & $1.113$ & $1.144$ & $1.200$ & $1.491$ &  &  &  &  &  &  &  \\
$k{=}5$ & $1.243$ & $1.155$ & $1.337$ & $1.221$ & $1.996$ &  &  &  &  &  &  \\
$k{=}6$ & $1.176$ & $1.161$ & $1.267$ & $1.264$ & $1.279$ & $2.289$ &  &  &  &  &  \\
$k{=}7$ & $1.207$ & $1.155$ & $1.313$ & $1.244$ & $1.468$ & $1.380$ & $2.949$ &  &  &  &  \\
$k{=}8$ & $1.170$ & $1.149$ & $1.294$ & $1.262$ & $1.310$ & $1.482$ & $0.9304$ & $4.875$ &  &  &  \\
$k{=}9$ & $1.125$ & $1.111$ & $1.269$ & $1.141$ & $1.162$ & $1.215$ & $0.5992$ & $2.994$ & $8.493$ &  &  \\
$k{=}10$ & $1.148$ & $1.120$ & $1.290$ & $1.175$ & $1.192$ & $1.247$ & $0.6686$ & $2.496$ & $-0.4112$ & $16.37$ &  \\
$k{=}11$ & $1.202$ & $1.137$ & $1.420$ & $1.288$ & $1.457$ & $1.207$ & $1.562$ & $0.9683$ & $6.004$ & $-0.1832$ & $72.46$ \\
\bottomrule
\end{tabular}}

\vspace{2ex}

\centerline{\textbf{Pantheon+ (flux-avg.)}}
\smallskip

\resizebox{\textwidth}{!}{
\begin{tabular}{l|ccccccccccc}
\toprule
 & $k{=}1$ & $k{=}2$ & $k{=}3$ & $k{=}4$ & $k{=}5$ & $k{=}6$ & $k{=}7$ & $k{=}8$ & $k{=}9$ & $k{=}10$ & $k{=}11$ \\
 & $z{=}0.05$ & $z{=}0.15$ & $z{=}0.30$ & $z{=}0.40$ & $z{=}0.50$ & $z{=}0.60$ & $z{=}0.70$ & $z{=}0.80$ & $z{=}1.00$ & $z{=}1.60$ & $z{=}2.27$ \\
\midrule
$10^{2} \log r_p$ & $1.1086$ & $0.53686$ & $0.27506$ & $1.3596$ & $-0.67832$ & $-1.7122$ & $-0.78229$ & $-4.5276$ & $2.8494$ & $-6.4285$ & $2.5618$ \\
\midrule
$10^{4} \Sigma_{kl},\,k{=}1$ & $2.203$ &  &  &  &  &  &  &  &  &  &  \\
$k{=}2$ & $1.927$ & $2.379$ &  &  &  &  &  &  &  &  &  \\
$k{=}3$ & $2.063$ & $2.226$ & $2.744$ &  &  &  &  &  &  &  &  \\
$k{=}4$ & $2.003$ & $2.143$ & $2.245$ & $2.826$ &  &  &  &  &  &  &  \\
$k{=}5$ & $2.132$ & $2.206$ & $2.546$ & $2.445$ & $3.600$ &  &  &  &  &  &  \\
$k{=}6$ & $2.107$ & $2.328$ & $2.607$ & $2.603$ & $2.814$ & $4.528$ &  &  &  &  &  \\
$k{=}7$ & $2.062$ & $2.276$ & $2.673$ & $2.477$ & $3.103$ & $3.373$ & $5.797$ &  &  &  &  \\
$k{=}8$ & $2.042$ & $2.264$ & $2.528$ & $2.532$ & $2.609$ & $3.273$ & $2.154$ & $8.430$ &  &  &  \\
$k{=}9$ & $2.058$ & $2.414$ & $2.695$ & $2.409$ & $2.423$ & $2.698$ & $1.646$ & $5.699$ & $13.68$ &  &  \\
$k{=}10$ & $2.590$ & $2.804$ & $3.190$ & $3.309$ & $3.230$ & $3.739$ & $2.636$ & $5.958$ & $2.855$ & $27.70$ &  \\
$k{=}11$ & $3.061$ & $3.135$ & $3.548$ & $4.081$ & $3.919$ & $3.552$ & $3.729$ & $3.636$ & $10.16$ & $5.728$ & $86.58$ \\
\bottomrule
\end{tabular}}

\end{table}

\begin{table}[p]
\centering
\caption{Compressed $\log r_p$ data product for DES-Dovekie.}
\label{tab:logrp_DESY5}
\setlength{\tabcolsep}{3pt}

\centerline{\textbf{DES-Dovekie (not flux-avg.)}}
\smallskip

\resizebox{\textwidth}{!}{
\begin{tabular}{l|ccccccccccc}
\toprule
 & $k{=}1$ & $k{=}2$ & $k{=}3$ & $k{=}4$ & $k{=}5$ & $k{=}6$ & $k{=}7$ & $k{=}8$ & $k{=}9$ & $k{=}10$ & $k{=}11$ \\
 & $z{=}0.05$ & $z{=}0.15$ & $z{=}0.30$ & $z{=}0.40$ & $z{=}0.50$ & $z{=}0.60$ & $z{=}0.70$ & $z{=}0.80$ & $z{=}0.90$ & $z{=}1.00$ & $z{=}1.15$ \\
\midrule
$10^{2} \log r_p$ & $-0.59803$ & $-2.3717$ & $-1.8972$ & $-2.1900$ & $-2.1823$ & $-1.9003$ & $-1.2380$ & $-2.2045$ & $-2.6148$ & $-3.5809$ & $-6.8597$ \\
\midrule
$10^{4} \Sigma_{kl},\,k{=}1$ & $3.398$ &  &  &  &  &  &  &  &  &  &  \\
$k{=}2$ & $3.088$ & $3.586$ &  &  &  &  &  &  &  &  &  \\
$k{=}3$ & $3.157$ & $2.943$ & $3.404$ &  &  &  &  &  &  &  &  \\
$k{=}4$ & $2.915$ & $2.962$ & $2.946$ & $3.096$ &  &  &  &  &  &  &  \\
$k{=}5$ & $3.070$ & $2.946$ & $3.139$ & $2.911$ & $3.339$ &  &  &  &  &  &  \\
$k{=}6$ & $3.014$ & $2.977$ & $3.041$ & $2.911$ & $2.975$ & $3.236$ &  &  &  &  &  \\
$k{=}7$ & $3.087$ & $2.977$ & $3.112$ & $2.890$ & $3.117$ & $2.977$ & $3.486$ &  &  &  &  \\
$k{=}8$ & $3.037$ & $2.998$ & $3.078$ & $2.916$ & $3.018$ & $3.068$ & $3.069$ & $3.787$ &  &  &  \\
$k{=}9$ & $3.020$ & $2.954$ & $3.115$ & $2.943$ & $3.132$ & $3.013$ & $3.155$ & $3.052$ & $5.364$ &  &  \\
$k{=}10$ & $3.067$ & $3.004$ & $3.086$ & $2.914$ & $3.067$ & $3.084$ & $3.150$ & $3.195$ & $3.062$ & $5.964$ &  \\
$k{=}11$ & $2.969$ & $2.935$ & $3.142$ & $2.969$ & $3.176$ & $3.039$ & $3.161$ & $3.086$ & $3.973$ & $1.394$ & $21.73$ \\
\bottomrule
\end{tabular}}

\vspace{2ex}

\centerline{\textbf{DES-Dovekie (flux-avg.)}}
\smallskip

\resizebox{\textwidth}{!}{
\begin{tabular}{l|ccccccccccc}
\toprule
 & $k{=}1$ & $k{=}2$ & $k{=}3$ & $k{=}4$ & $k{=}5$ & $k{=}6$ & $k{=}7$ & $k{=}8$ & $k{=}9$ & $k{=}10$ & $k{=}11$ \\
 & $z{=}0.05$ & $z{=}0.15$ & $z{=}0.30$ & $z{=}0.40$ & $z{=}0.50$ & $z{=}0.60$ & $z{=}0.70$ & $z{=}0.80$ & $z{=}0.90$ & $z{=}1.00$ & $z{=}1.15$ \\
\midrule
$10^{2} \log r_p$ & $0.17156$ & $-2.3919$ & $-1.4898$ & $-1.7585$ & $-1.0210$ & $-1.6659$ & $-0.40884$ & $-1.9147$ & $-0.62430$ & $-2.0344$ & $-6.1991$ \\
\midrule
$10^{4} \Sigma_{kl},\,k{=}1$ & $5.108$ &  &  &  &  &  &  &  &  &  &  \\
$k{=}2$ & $4.239$ & $4.883$ &  &  &  &  &  &  &  &  &  \\
$k{=}3$ & $4.339$ & $4.011$ & $4.653$ &  &  &  &  &  &  &  &  \\
$k{=}4$ & $3.853$ & $4.008$ & $3.951$ & $4.244$ &  &  &  &  &  &  &  \\
$k{=}5$ & $4.264$ & $4.059$ & $4.251$ & $4.024$ & $4.778$ &  &  &  &  &  &  \\
$k{=}6$ & $4.320$ & $4.039$ & $4.110$ & $3.948$ & $4.251$ & $4.644$ &  &  &  &  &  \\
$k{=}7$ & $4.358$ & $4.060$ & $4.206$ & $3.902$ & $4.365$ & $4.257$ & $4.896$ &  &  &  &  \\
$k{=}8$ & $4.274$ & $4.072$ & $4.189$ & $3.964$ & $4.284$ & $4.400$ & $4.391$ & $5.406$ &  &  &  \\
$k{=}9$ & $4.241$ & $3.972$ & $4.082$ & $3.966$ & $4.407$ & $4.414$ & $4.511$ & $4.351$ & $7.760$ &  &  \\
$k{=}10$ & $4.599$ & $4.175$ & $4.226$ & $3.976$ & $4.525$ & $4.657$ & $4.686$ & $4.801$ & $4.596$ & $8.518$ &  \\
$k{=}11$ & $4.089$ & $3.769$ & $4.199$ & $3.975$ & $4.481$ & $4.549$ & $4.644$ & $4.612$ & $5.847$ & $3.249$ & $27.76$ \\
\bottomrule
\end{tabular}}

\end{table}

\begin{table}[p]
\centering
\caption{Compressed $\log r_p$ data product for Union3 (pre-binned at the catalog level so we don't apply flux averaging).}
\label{tab:logrp_Union3}
\setlength{\tabcolsep}{3pt}

\centerline{\textbf{Union3 (not flux-avg.)}}
\smallskip

\resizebox{\textwidth}{!}{
\begin{tabular}{l|ccccccccccc}
\toprule
 & $k{=}1$ & $k{=}2$ & $k{=}3$ & $k{=}4$ & $k{=}5$ & $k{=}6$ & $k{=}7$ & $k{=}8$ & $k{=}9$ & $k{=}10$ & $k{=}11$ \\
 & $z{=}0.05$ & $z{=}0.15$ & $z{=}0.30$ & $z{=}0.40$ & $z{=}0.50$ & $z{=}0.60$ & $z{=}0.70$ & $z{=}0.80$ & $z{=}1.00$ & $z{=}1.60$ & $z{=}2.27$ \\
\midrule
$10^{2} \log r_p$ & $3.8188$ & $3.5421$ & $1.7181$ & $2.7186$ & $1.8590$ & $2.2943$ & $-0.16623$ & $-0.03631$ & $4.1078$ & $-2.6234$ & $-6.7699$ \\
\midrule
$10^{4} \Sigma_{kl},\,k{=}1$ & $15.79$ &  &  &  &  &  &  &  &  &  &  \\
$k{=}2$ & $18.43$ & $22.28$ &  &  &  &  &  &  &  &  &  \\
$k{=}3$ & $18.79$ & $22.48$ & $23.20$ &  &  &  &  &  &  &  &  \\
$k{=}4$ & $17.95$ & $21.72$ & $22.17$ & $21.70$ &  &  &  &  &  &  &  \\
$k{=}5$ & $18.61$ & $22.35$ & $22.91$ & $22.07$ & $23.34$ &  &  &  &  &  &  \\
$k{=}6$ & $18.27$ & $22.04$ & $22.57$ & $21.82$ & $22.51$ & $22.96$ &  &  &  &  &  \\
$k{=}7$ & $18.44$ & $22.20$ & $22.79$ & $21.95$ & $22.80$ & $22.56$ & $24.01$ &  &  &  &  \\
$k{=}8$ & $18.42$ & $22.20$ & $22.79$ & $22.00$ & $22.82$ & $22.67$ & $23.17$ & $24.20$ &  &  &  \\
$k{=}9$ & $18.43$ & $22.20$ & $22.75$ & $21.96$ & $22.85$ & $22.55$ & $22.92$ & $23.03$ & $25.91$ &  &  \\
$k{=}10$ & $18.40$ & $22.19$ & $22.75$ & $22.00$ & $22.72$ & $22.66$ & $22.47$ & $23.81$ & $21.12$ & $39.84$ &  \\
$k{=}11$ & $18.53$ & $22.32$ & $22.99$ & $22.18$ & $23.19$ & $22.57$ & $23.84$ & $21.99$ & $28.01$ & $17.63$ & $230.5$ \\
\bottomrule
\end{tabular}}

\end{table}


\bibliographystyle{JHEP}
\bibliography{main}

\end{document}